\UseRawInputEncoding
\documentclass[prl,twocolumn,groupedaddress,floatfix,footinbib]{revtex4-1}

\usepackage{graphicx}
\usepackage{bm}
\usepackage{amsmath}
\usepackage{amsfonts} 
\usepackage{amssymb}
\usepackage{latexsym}
\usepackage{color}
\usepackage{mathrsfs}
\usepackage{braket}
\usepackage{enumerate}
\usepackage[usenames,dvipsnames]{xcolor}
\usepackage[colorlinks=true,citecolor=blue,linkcolor=blue,urlcolor=blue,backref=true]{hyperref}
\usepackage{float}
\usepackage{mathtools}
\usepackage[inline]{enumitem}
\usepackage{upgreek}
\usepackage{accents}

\begin{document}

\title{Uniqueness of the Phase Transition in Many-Dipole Cavity Quantum Electrodynamical Systems}
\author{Adam Stokes}\email{adamstokes8@gmail.com}
\affiliation{Department of Physics and Astronomy, University of Manchester, Oxford Road, Manchester M13 9PL, United Kingdom}
\author{Ahsan Nazir}\email{ahsan.nazir@manchester.ac.uk}
\affiliation{Department of Physics and Astronomy, University of Manchester, Oxford Road, Manchester M13 9PL, United Kingdom}

\date{\today}

\begin{abstract}
The possibility of a superradiant phase transition in light-matter systems is the subject of much debate, due to numerous apparently conflicting no-go and counter no-go theorems. 
Using an arbitrary-gauge approach we show 
that a {\em unique} phase transition does 
occur in archetypal many-dipole cavity QED systems, and that it manifests unambiguously via a macroscopic gauge-invariant polarisation. We 
find that the gauge choice controls the extent to which this polarisation is included as part of the radiative quantum subsystem and thereby determines the degree to which the abnormal phase is classed as superradiant. This resolves the long-standing paradox of 
no-go and counter no-go theorems for superradiance, which are shown to refer to different definitions of radiation. 
\end{abstract}

\maketitle
  

Superradiance was originally described by Dicke \cite{dicke_coherence_1954}, and since then it has 
received a great deal of attention (see e.g.~\cite{kirton_introduction_2019} for a recent introduction). A superradiant phase of a light-matter system is one in which a macroscopic number of photons arises due to the interaction between many dipoles. The possibility of such a phase transition within the Dicke model was first recognised some time ago \cite{hepp_superradiant_1973,wang_phase_1973}. Later, seminal contributions were made in the connection with quantum chaos 
\cite{holstein_field_1940,emary_chaos_2003,emary_quantum_2003}. The topic now includes extended Dicke models \cite{carmichael_higher_1973,hioe_phase_1973,pimentel_phase_1975,emeljanov_appearance_1976,sung_phase_1979}, driven and open systems, semi-classical descriptions \cite{grimsmo_dissipative_2013,klinder_dynamical_2015,gegg_superradiant_2018,kirton_superradiant_2018,peng_unified_2019}, and artificial QED systems \cite{yamanoi_polariton_1979,haug_quantum_1994,lee_first-order_2004,nataf_no-go_2010,viehmann_superradiant_2011,todorov_intersubband_2012,leib_synchronized_2014,bamba_stability_2014,bamba_superradiant_2016,bamba_circuit_2017,jaako_ultrastrong-coupling_2016,de_bernardis_cavity_2018}.

One of the most controversial aspects of theoretical studies has been the validity of so-called ``no-go theorems", which prohibit a superradiant phase and are proved in the Coulomb-gauge. The original no-go theorem  \cite{rzazewski_phase_1975} actually prohibits a phase transition of any kind, but neglects direct electrostatic interactions, whose presence are a defining feature of a correct Coulomb-gauge model. This theorem, and variants thereof, have been both refuted and confirmed in numerous subsequent works \cite{kudenko_interatomic_1975,rzazewski_remark_1976,emeljanov_appearance_1976,knight_are_1978,bialynicki-birula_no-go_1979,yamanoi_polariton_1979,sung_phase_1979,rzazewski_stability_1991,keeling_coulomb_2007,nataf_no-go_2010,vukics_adequacy_2012,vukics_elimination_2014,bamba_stability_2014,tufarelli_signatures_2015,grieser_depolarization_2016,andolina_cavity_2019}. It has been suggested that where natural atomic systems admit a no-go theorem certain artificial atomic systems do not \cite{nataf_no-go_2010} (though see also \cite{viehmann_superradiant_2011,PhysRevLett.109.179301}). However, in the multipolar-gauge the superradiant phase transition also appears to be automatically recovered for conventional cavity QED systems \cite{vukics_adequacy_2012,vukics_elimination_2014}. 

Further permutations of these results are available. For example,
if explicit dipole-dipole interactions that are not naturally present are added into the multipolar-gauge description, then a no-go theorem re-emerges \cite{bialynicki-birula_no-go_1979,haug_quantum_1994,de_bernardis_breakdown_2018,stefano_resolution_2019}. A very recent contribution \cite{andolina_cavity_2019} argues 
without the two-level approximation that a superradiant phase is impossible, but this treatment considers only the radiative quantum subsystem and is again proved in the Coulomb gauge. If, rather than just the radiative subsystem, one also considers variations in the electrostatic interactions that are present within the Coulomb-gauge, then an apparently different ferroelectric phase transition is predicted. This, however, does not lead to superradiance \cite{keeling_coulomb_2007}. 
Thus, despite numerous contributions spanning several decades, the occurrence and nature of the phase transition in generic many-emitter light-matter systems, and how this relates to the choice of gauge, are fundamental questions whose answers remain unclear,
yet still 
highly relevant~\cite{PhysRevLett.122.017401,andolina_cavity_2019,PhysRevLett.123.207402,guerci2020superradiant,andolina2020theory}.

Here we resolve these fundamental 
issues 
by proving that a {\em unique} physical phase transition does 
occur in generic many-dipole {\em cavity} QED systems and that the abnormal phase of the system is unambiguously signalled by a macroscopic average of the gauge-invariant transverse polarisation field ${\bf P}_{\rm T}$. 
This equals the longitudinal electric field ${\bf E}_{\rm L}$ except at the point-dipole positions themselves. Crucial to the resolution provided 
is the recognition that QED subsystems are gauge-relative, meaning that each gauge provides different {\emph{gauge-invariant}} definitions of the light and matter subsystems. 
Whether the abnormal phase is characterised as ferroelectric or as superradiant depends on the extent to which ${\bf E}_{\rm L}={\bf P}_{\rm T}$ is included within the radiative quantum subsystem and this is controlled by the gauge choice. We thereby show that the different viewpoints provided by different gauges are not contradictory, but in fact equivalent, as required. In particular, correct no-go statements such as in Ref.~\cite{andolina_cavity_2019} are reconciled with correct counter no-go statements such as in Ref.~\cite{vukics_adequacy_2012}. 
Such results are found to be different ways of viewing the same phenomenon in terms of physically different quantum subsystems. By converting the apparent gauge non-invariance of the phase transition into a proof of gauge-invariance, our results resolve 
the associated long-standing controversies.  

A related but separate point is that level truncation of material dipoles causes a breakdown of gauge-invariance \cite{de_bernardis_breakdown_2018,stokes_gauge_2019,stefano_resolution_2019,roth_optimal_2019}. 
Using numerical results for finite numbers of dipoles we show that accurate two-level model predictions can be identified. It is reasonable to conclude that the same two-level truncation will be accurate in the thermodynamic limit. Thus, arbitrary-gauge QED 
is also 
capable of eliminating any further quantitative ambiguity resulting from the use of material two-level truncation.


We begin by deriving an arbitrary gauge Dicke Hamiltonian. 
We adopt a general formulation of QED in which the gauge is selected by a real parameter $\alpha$. We consider $N$ identical electric dipoles each 
described by a classical centre-of-mass position ${\bf R}_\mu$ and 
a dipole moment operator ${\hat {\bf d}}_\mu=-e{\bf r}_\mu$. The dipoles interact with a common electromagnetic field described by transverse-electric and magnetic fields ${\bf E}_{\rm T}$ and ${\bf B}$ respectively. We 
obtain the Hamiltonian for the system 
from first principles~\footnote{Please see the Supplemental Material, which includes Refs.~\cite{stokes_master_2018} and~\cite{williamson_algebraic_1936}.}, 
which can be written in the gauge-invariant form \cite{stokes_gauge_2019}
$H = E_{\rm matter}+E_{\rm field}$, 
where
$E_{\rm matter} := \sum_{\mu =1}^N  {1\over 2}m {\dot {\bf r}}_\mu^2 +V +V_{\rm dip}$ and 
$E_{\rm field}:={1\over 2}\int d^3 x\, \left[{\bf E}_{\rm T}({\bf x})^2+{\bf B}({\bf x})^2\right]$. 
Here $V$ denotes the total intra-dipole potential, and $V_{\rm dip}$ denotes the inter-dipole electrostatic energy. The $\alpha$-dependent canonical momenta are found to be 
\begin{align}
{\bf p}_{\mu\alpha} &= m{\dot {\bf r}}_\mu - e(1-\alpha){\bf A}({\bf R}_\mu),\label{mom1}\\
{\bf \Pi}_{\alpha}({\bf x}) &= -{\bf E}_{\rm T}({\bf x}) - {\bf P}_{{\rm T}\alpha}({\bf x}),\label{mom2}
\end{align}
where ${\bf A}$ is the gauge-invariant transverse vector potential such that ${\bf E}_{\rm T}=-{\dot {\bf A}}$ and ${\bf P}_{{\rm T}\alpha}$ is the $\alpha$-gauge transverse polarisation given by
${\bf P}_{{\rm T}\alpha}({\bf x}) = \alpha {\bf P}_{{\rm T}}({\bf x})$, 
with ${\bf P}_{\rm T}({\bf x}) =\sum_{\mu =1}^N{\hat {\bf d}}_\mu\cdot \delta^{\rm T}({\bf x}-{\bf R}_\mu)$. The canonical commutation relations are $[r_{\mu,i},p_{\nu,j}]=i\delta_{\mu\nu}\delta_{ij}$ and $[A_{{\rm T},i}({\bf x}),\Pi_{{\rm T},j}({\bf x}')]=\delta_{ij}^{\rm T}({\bf x}-{\bf x}')$. All other commutators between canonical operators vanish. 
The canonical momenta of different gauges are unitarily related via ${\bf X}_\alpha= R_{\alpha\alpha'}{\bf X}_{\alpha'}R_{\alpha'\alpha}$, where ${\bf X}={\bf p},\,{\bf \Pi}$ and
$R_{\alpha\alpha'} = \exp\left(i(\alpha-\alpha')\sum_{\mu=1}^N {\hat {\bf d}}_\mu \cdot {\bf A}\right)$.

We now restrict our attention to a single cavity mode with volume $v$, frequency $\omega$ 
and unit polarisation vector ${\bm \varepsilon}$, 
described by bosonic operators $a_\alpha,~a_\alpha^\dagger$ with $[a_\alpha,a_\alpha^\dagger]=1$.  
The restriction is imposed consistently on all fields including the transverse delta-function $\delta^{\rm T}$. This eliminates the need to regularise ${\bf P}_{\rm T}$ \cite{vukics_fundamental_2015}, and ensures that the transverse commutation relation for the canonical fields is preserved. The fundamental kinematic relations given by Eqs.~(\ref{mom1}) and (\ref{mom2}) are therefore also preserved. In order to obtain a Dicke Hamiltonian we next 
take the limit of closely spaced dipoles around the origin; ${\bf R}_\mu \approx {\bf 0}$ 
and we approximate the dipoles as two-level systems. Further details of all approximations used are given in~\cite{Note1}.  
We introduce the collective operators
$J_\alpha^i = \sum_{\mu=1}^N \sigma^i_{\mu\alpha}$, with $i=\pm,z$,
where $\sigma^\pm_{\mu\alpha}$ are the raising and lowering operators of the $\mu$'th two-level dipole and $\sigma^z_{\mu\alpha} = [\sigma^+_{\mu\alpha},\sigma^-_{\mu\alpha}]/2$. 
We also introduce cavity bosonic operators $c_\alpha$ and $c_\alpha^\dagger$, which incorporate both the bare cavity energy and the ${\bf A}^2$-term that results when Eq.~(\ref{mom1}) is substituted into the energy $H$ [Supplementary Eqs.~(72), (73)]. 
The resulting arbitrary-gauge Dicke 
Hamiltonian is
\begin{align}\label{h2}
H^{\alpha,2} =&\, \omega_m J_\alpha^z +{N\over 2}(\epsilon_0+\epsilon_1)+{1\over 2} \rho d^2 +\omega_\alpha \left(c_\alpha^\dagger c_\alpha+{1\over 2}\right)  \nonumber \\ &-{{\cal C}_\alpha\over N}\left(J_\alpha^++J_\alpha^-\right)^2 
-i{g_\alpha'\over \sqrt{N}} (J_\alpha^+-J_\alpha^-)(c_\alpha^\dagger +c_\alpha) \nonumber \\ &+i{g_\alpha\over \sqrt{N}} (J_\alpha^++J_\alpha^-)(c_\alpha^\dagger -c_\alpha),
\end{align}
where $\omega_m=\epsilon_1-\epsilon_0$, 
${\omega_\alpha}^2 = \omega^2 + e^2(1-\alpha)^2\rho/m$,
${\cal C}_\alpha := \rho d^2 (1-\alpha^2)/2$, 
$g_\alpha':=(1-\alpha)\omega_m d \sqrt{\rho/(2\omega_\alpha)}$, 
and $g_\alpha:=\alpha d \sqrt{\rho \omega_\alpha /2}$,
with $d :={\bm \varepsilon}\cdot {\bf d}$. Here $\rho =N/v$ remains finite in the thermodynamic limit $N\to \infty$, $v\to \infty$. 
Although the non-truncated Hamiltonian $H$ is unique, we now have a continuous infinity of Dicke Hamiltonians $H^{\alpha,2}$ such that $H^{\alpha,2}$ and $H^{\alpha',2}$ are not equal when $\alpha\neq \alpha'$ \cite{de_bernardis_breakdown_2018,stokes_gauge_2019,stefano_resolution_2019,roth_optimal_2019}. 
This breaking of gauge-invariance will turn out not to be a barrier to eliminating all ambiguities regarding the occurrence and nature of a quantum phase transition.

To take the thermodynamic limit we use a Holstein-Primakoff map defined by
$J_\alpha^z = b_\alpha^\dagger b_\alpha -{N/ 2}$, $J_\alpha^+ = b_\alpha^\dagger \sqrt{N-b_\alpha^\dagger b_\alpha}$, and $J_\alpha^-=(J_\alpha^+)^\dagger$,
where $[b_\alpha,b_\alpha^\dagger]=1$ \cite{holstein_field_1940,emary_chaos_2003,emary_quantum_2003}. The Hamiltonian obtained by substituting these expressions into Eq.~(\ref{h2}) is denoted $H_{\rm th}^{\alpha,2}$. 
We first consider the material part of $H_{\rm th}^{\alpha,2}$, which can be written $H_{{\rm th,}m}^{\alpha,2} ={\tilde \omega}_m^\alpha l^\dagger_\alpha l_\alpha+{1\over 2}({\tilde \omega}_m^\alpha -\omega_m)$ where $[l_\alpha, l_\alpha^\dagger]=1$ and
\begin{align}
{{\tilde \omega}_m^\alpha}^2 &= \omega_m( \omega_m- 4 {\cal C}_\alpha). \label{wmtil}
\end{align}
The mode operators $l_\alpha,~l_\alpha^\dagger$ are related to $b_\alpha$ and $b_\alpha^\dagger$ by a local Bogoliubov transformation that incorporates the contribution $V_{\rm dip}$ (see Eqs.~(79) and (80) in~\cite{Note1}). This results in the renormalised frequency ${\tilde \omega}_m^\alpha$ in Eq. (\ref{wmtil}). Reality of ${\tilde \omega}_m^\alpha$ requires that
\begin{align}\label{ferro}
\omega_m\geq {4 {\cal C}_\alpha}=2\rho d^2 (1-\alpha^2).
\end{align}
When the electrostatic interaction strength ${\cal C}_\alpha$ is large enough this inequality may be violated signalling a phase transition. We refer to this transition as ferroelectric, because it is completely independent of the radiative mode. Inequality (\ref{ferro}) generalises the result of Keeling obtained when $\alpha=0$ (Coulomb gauge)~\cite{keeling_coulomb_2007}. Violation of inequality (\ref{ferro}) cannot occur in the multipolar gauge $\alpha=1$, which does not therefore admit a purely ferroelectric phase. In what follows this finding will be reconciled with our claim that a unique phase transition is predicted within all gauges. We show further that only in the Coulomb gauge does the phase transition appear purely ferroelectric.

We now consider the thermodynamic-limit of the total Hamiltonian, which is~\cite{Note1} 
\begin{align}\label{norm}
H_{\rm th}^{\alpha,2,\rm i} =& \,E^{\rm i}_{\alpha+} {f^{\rm i}_\alpha}^\dagger f^{\rm i}_\alpha + E^{\rm i}_{\alpha-}  {c^{\rm i}_\alpha}^\dagger c^{\rm i}_\alpha +{1\over 2}( E^{\rm i}_{\alpha+} +E^{\rm i}_{\alpha-} )+C^{\rm i}
\end{align}
where the superscript $\cdot^{\rm i}$ is either ${\rm i}={\rm n}$ for normal-phase, or ${\rm i}={\rm a}$ for abnormal-phase. 
The polariton operators $f^{\rm i}_{\alpha},\,c^{\rm i}_\alpha$ are bosonic satisfying $[f^{\rm i}_{\alpha},{f^{\rm i}_{\alpha}}^\dagger]=1=[c^{\rm i}_{\alpha},{c^{\rm i}_{\alpha}}^\dagger]$ with all other commutators 
vanishing. In the normal phase, ${\rm i}={\rm n}$, the zero-point constant in Eq.~(\ref{norm}) is $C^{\rm n} = N\epsilon_0+\left(\rho d^2-\omega_m\right)/2$ and the polariton energies are 
\begin{align}
2{E^{\rm n}_{\alpha\pm}}^2 =&\, 8{\tilde g}_\alpha {\tilde g}_\alpha'+ {{\tilde \omega}_m^\alpha}^2 + \omega_\alpha^2 \pm \sqrt \bigg(\left[{{\tilde \omega}_m^\alpha}^2 - \omega_\alpha^2 \right]^2 \nonumber \\ & +16[{\tilde \omega}_m^\alpha  {\tilde g}_\alpha' + \omega_\alpha  {\tilde g}_\alpha] [{\tilde \omega}_m^\alpha {\tilde g}_\alpha+\omega_\alpha {\tilde g}_\alpha'] \bigg),
\end{align}
where
${\tilde g}_\alpha= g_\alpha\sqrt{\omega_m /{\tilde \omega}_m^\alpha}$ and 
${\tilde g}'_\alpha= g_\alpha'\sqrt{{\tilde \omega}_m^\alpha / \omega_m}$. 
The coupling strength at which the lower polariton energy $E^{\rm n}_{\alpha-}$ is no longer real signals the onset of the abnormal phase and the breakdown of $H_{\rm th}^{\alpha,2,{\rm n}}$. Reality of $E^{\rm n}_{\alpha-}$ requires that
\begin{align}\label{con1}
\omega_m \left(\omega_m -2 \rho d^2\right) \left(\omega_\alpha^2 -2\omega_m \rho d^2(1-\alpha)^2\right) \geq 0.
\end{align}
From the Thomas-Reiche-Kuhn (TRK) inequality
${e^2/ m} \geq 2\omega_m d^2$ 
it follows that $\omega_\alpha^2 \geq  2 \omega_m \rho d^2 (1-\alpha)^2$. Therefore, by inequality (\ref{con1}) $E_{\alpha-}$ is real if and only if
\begin{align}\label{pt}
\omega_m \geq 2\rho d^2.
\end{align}
This simple gauge-invariant result defines the normal phase. Inequality (\ref{pt}) is stronger than inequality (\ref{ferro}), so ${\tilde \omega}_m^\alpha$ in Eq. (\ref{wmtil}) is also real when inequality (\ref{pt}) is satisfied.



The Hamiltonian $H_{\rm th}^{\alpha,2,\rm a}$ for the abnormal phase takes over from $H_{\rm th}^{\alpha,2,\rm n}$ when inequality (\ref{pt}) is violated. It is obtained as the thermodynamic limit of $H^{\alpha,2}$ written, via the Holstein-Primakoff map, in terms of displaced modes $f_\alpha$ and $c'_\alpha$ such that $b_\alpha= f_\alpha -\sqrt{\beta_\alpha}$ and $c_\alpha=c'_\alpha+i\sqrt{\gamma_\alpha}$, where 
$\beta_\alpha = \beta:=N(1-\tau)/2$ 
and $\gamma_\alpha = {N g_\alpha^2}\left(1-\tau^2\right)/\omega_\alpha^2$ are of order $N$, 
with
$\tau = 
{\omega_m /(2\rho d^2)}$.
Note that $\beta_\alpha=\beta$ is $\alpha$-independent indicating that 
the ``material" mode is always displaced by the same macroscopic quantity. 
On the other hand, $\gamma_\alpha$ is $\alpha$-dependent, so the extent to which the ``radiative" mode is displaced depends on the chosen definition of radiation. In particular, $\gamma_0=0$, so in the Coulomb gauge {\em only} the material mode is displaced.  
In the abnormal-phase, ${\rm i}={\rm a}$, the zero-point constant in Eq.~(\ref{norm}) is $C^{\rm a}=N\left[\epsilon_0-{\omega_m}(1-\tau)^2\right]/(4\tau)-\rho d^2/2$ and the polariton energies are 
\begin{align}
2{E^{\rm a}_{\alpha\pm}}^2 =&\, 8\underaccent{\tilde}{g}_\alpha \underaccent{\tilde}{g}_\alpha'+ {\underaccent{\tilde}{\omega}_m^\alpha}^2 + \omega_\alpha^2 \pm \sqrt \bigg(\left[{\underaccent{\tilde}{\omega}_m^\alpha}^2 - \omega_\alpha^2 \right]^2 \nonumber \\ & +16[\underaccent{\tilde}{\omega}_m^\alpha  \underaccent{\tilde}{g}_\alpha' + \omega_\alpha  \underaccent{\tilde}{g}_\alpha'] [\underaccent{\tilde}{\omega}_m^\alpha \underaccent{\tilde}{g}_\alpha+\omega_\alpha \underaccent{\tilde}{g}_\alpha'] \bigg), \label{apol}
\end{align}
where
${\underaccent{\tilde}{\omega}_m^\alpha}^2 = {{\omega_m}^2}\left[1-(1-\alpha^2)\tau^2\right]/\tau^2$, while 
$\underaccent{\tilde}{g}'_\alpha = g_\alpha'\sqrt{\tau\underaccent{\tilde}{\omega}_m^\alpha /\omega_m}$, 
and $\underaccent{\tilde}{g}_\alpha = g_\alpha\sqrt{\tau \omega_m / \underaccent{\tilde}{\omega}_m^\alpha}$.
The material frequency $\underaccent{\tilde}{\omega}_m^\alpha$ is real provided $(2\rho d^2)^2\geq {\omega_m}^2(1-\alpha^2)$ and the lower polariton energy $E_{\alpha-}^{\rm a}$ is real provided
\begin{align}\label{ab}
\left([2 \rho d^2]^2-{\omega_m}^2\right) \left(\omega_\alpha^2 -{\omega_m}^2(1-\alpha)^2\right) \geq 0.
\end{align}
In the abnormal phase we have $2\rho d^2 \geq \omega_m$ implying that $\omega_\alpha^2 -{\omega_m}^2(1-\alpha)^2\geq 0$ and therefore 
that $E_{\alpha-}^{\rm a}$ is real. 
At the critical coupling point where $2\rho d^2=\omega_m$ the Hamiltonians $H_{\rm th}^{\alpha,2,\rm n}$ and $H_{\rm th}^{\alpha,2,\rm a}$ coincide. 
We have therefore obtained a description of the thermodynamic limit for all coupling strengths. The polariton energies constitute different two-level approximated results in each different gauge $\alpha$. This is shown in Supplementary Figure 1. However, every gauge's approximate (Dicke) model predicts exactly one ground state phase transition occurring when $\omega_m=2\rho d^2$ and the ground state is unique within the non-truncated theory. Therefore, inequality (\ref{pt}) should be interpreted as predicting a unique phase transition. 
%
%
%
However, the nature of the phase transition appears to be different depending on the value of $\alpha$. In the Coulomb gauge for example, 
it is necessarily purely ferroelectric, whereas this is impossible in the multipolar-gauge. 

The radiative classification of a unique phase transition 
will naturally depend on the definition of radiation and the latter is controlled by the gauge. Evidently, the {\em subsystem gauge-relativity} of QED \cite{stokes_ultrastrong_2019}, is strongly exemplified by the phase transition phenomenon. 
To understand the physical meaning of ``matter" and ``radiation" in the gauge $\alpha$ 
we note that 
the {\em total} multipolar polarisation of $N$ dipoles is ${\bf P}({\bf x})= \sum_{\mu=1}^N{\bf d}_\mu \delta({\bf x}-{\bf R}_\mu)$. Since $\nabla\cdot {\bf E}=\rho=-\nabla \cdot {\bf P}$ it follows that 
for ${\bf x}\neq {\bf R}_\mu$ we have ${\bf P}_{\rm T}={\bf E}_{\rm L}$ and therefore ${\bf \Pi}_\alpha =-{\bf E}_{\rm T}-\alpha {\bf E}_{\rm L}$ [cf. Eq.~(\ref{mom2})]. Similarly, the material momentum ${\bf p}_\alpha$ of a dipole is given by Eq.~(\ref{mom1}) in which $q{\bf A}({\bf 0})$ is the electric dipole approximation (EDA) of $q{\bf A}({\bf r}) = {\bf P}_{\rm long}=\int d^3x {\bf E}_{\rm L\bf r}({\bf x})\times {\bf B}({\bf x})$, which is the momentum of the longitudinal field generated by $q$ at ${\bf r}$ with $\nabla\cdot {\bf E}_{\rm L\bf r}({\bf x})=q\delta({\bf x}-{\bf r})$.

As an example, one may consider the Coulomb-gauge in which ``matter" is fully dressed by ${\bf E}_{\rm L}$, i.e., ${\bf p}_0=m{\dot {\bf r}}+{\bf P}_{\rm long}$, so ``matter" as defined by ${\bf p}_0$ is not fully localised. Correspondingly, ``radiation" is defined using the field ${\bf \Pi}_0=-{\bf E}_{\rm T}$ alone. In the multipolar-gauge (within the EDA) matter is completely bare, i.e., ${\bf p}_1=m{\dot {\bf r}}$, and therefore fully localised. ``Radiation"  is correspondingly defined for ${\bf x}\neq {\bf 0}$ by the local (causal) total field ${\bf \Pi}_1=-{\bf D}_{\rm T}=-{\bf E}_{\rm T}-{\bf E}_{\rm L}=-{\bf E}$. More generally, $\alpha$ controls how the longitudinal electric degrees of freedom are shared out, thereby controlling the balance between localisation and electrostatic dressing in defining the quantum subsystem called ``matter". ``Radiation" is then defined using the canonical degrees of freedom left over. 

There are noteworthy gauges in between $\alpha=0$ and $\alpha=1$, such as gauges relative to which ground state ``virtual photons" are highly suppressed and for which the corresponding two-level model can sometimes offer a more accurate representation of the ground state than conventional quantum Rabi models \cite{stokes_gauge_2019} (see also~\cite{Note1}). What differs between gauges is the spacetime locatisation properties of ``material sources" and their dressing by virtual photons. In general the most operationally relevant definitions of the subsystems may depend on the available measurements, including their time- and length-scales. As a result, general statements about {\em measurable} photon condensation (superradiance), that are independent of experimental context, cannot be made. What can be demonstrated and is demonstrated below, is that there are no internal theoretical inconsistencies and no fundamental paradoxes. Previous no-go and counter no-go theorems refer to different definitions of radiation and so are not contradictory. They are in fact equivalent.

We now calculate the ground-state momentum ${\bf \Pi}_\alpha$ of radiation defined relative to gauge $\alpha$. 
This directly demonstrates strict equivalence of all gauges and reveals an unambiguous macroscopic manifestation of the abnormal phase. We allow the two-level truncation to be performed in an arbitrary gauge $\alpha'$. 
The $\alpha'$-gauge two-level approximation of an operator $o_\alpha = o_\alpha({\bf p}_\alpha,{\bf \Pi}_\alpha)$, denoted $o^{\alpha',2}_\alpha$, is found by expressing $o_\alpha$ in terms of $\alpha'$-gauge canonical operators followed by two-level truncation. For $\Pi_\alpha = {\bm \varepsilon}\cdot {\bf \Pi}_\alpha$ we have 
$\Pi_\alpha^{\alpha',2} = \Pi_{\alpha'}-{d}(\alpha-\alpha')(J_{\alpha'}^+ + J_{\alpha'}^-)/v$. We will see that in the thermodynamic limit the ground state value of $\Pi_\alpha^{\alpha',2}$ is actually independent of $\alpha'$, i.e. the prediction is gauge-invariant, so we return to the simpler notation $\Pi_{\rm \alpha,th}$. 
Using the Holstein-Primakoff representation, we find that 
$\Pi_{\alpha,\rm th}$ 
vanishes in the normal phase and in the abnormal phase is proportional to the identity. The calculation in~\cite{Note1} 
yields the simple result 
\begin{align}\label{abY}
\Pi_{\alpha,{\rm th}}^{\rm a} = \alpha \rho d\sqrt{1-\tau^2}={\alpha\over 2d}\sqrt{\left(2\rho d^2\right)^2-{\omega_m}^2}.
\end{align}
The factor of $\alpha$ in Eq.~(\ref{abY}) is highly significant. It demonstrates that the degree of superradiance in the abnormal phase is proportional to $\alpha$, 
with the minimum value of zero occurring only in the Coulomb-gauge.

To demonstrate equivalence between all gauges we calculate the $\alpha$-gauge transverse polarisation $P_{\rm T\alpha}=\alpha {\bm \varepsilon}\cdot{\bf P}_{\rm T}=\alpha(\Pi_0-\Pi_1)$, which is such that 
$P_{\rm T\alpha}^{\alpha',2} = \alpha {d\over v} (J_{\alpha'}^+ + J_{\alpha'}^-)$. This quantity is also $\alpha'$-independent in the thermodynamic limit.
In the normal phase $P_{\rm T\alpha, th}$ vanishes, whereas in the abnormal phase it is found by the same method that leads to Eq.~(\ref{abY}) to be
$P_{\rm T\alpha, th}^{\rm a}  =
- \alpha \rho d\sqrt{1-\tau^2}$. 
Eq. (\ref{abY}) then yields
$\Pi_{\alpha,{\rm th}}^{\rm a}=-P_{\rm T\alpha, th}^{\rm a}$, 
which since $-E_{\rm T, th}^{\rm a}=\Pi_{0,{\rm th}}^{\rm a}=0$, is seen to be nothing but the fundamental kinematic relation (\ref{mom2}). This establishes consistency between all gauges.  
The quantity $|P_{\rm T,th}^{\alpha',2,{\rm a}}|= |P_{\rm T\alpha, th}^{\alpha',2,{\rm a}}/\alpha|$ 
provides a gauge-invariant monotonic measure of the coupling-distance past the phase transition point. 
Thus, independent of the gauge the onset of the abnormal phase manifests in the form of a macroscopic value of the gauge-invariant field ${\bf P}_{\rm T}$;
\begin{align}\label{pabt}
P_{\rm T,th}^{\rm a}= - \rho d\sqrt{1-\tau^2}
\end{align}
which is plotted in Supplementary Figure 2. Within the present simplified Dicke-type treatment the field ${\bf P}_{\rm T}$ is independent of spatial position ${\bf x}$, but at a more fundamental level 
${\bf P}_{\rm T}$ coincides with the longitudinal electric field ${\bf E}_{\rm L}$ away from the dipole positions, i.e., for ${\bf x}\neq {\bf R}_\mu$. Whether one considers ${\bf E}_{\rm L}={\bf P}_{\rm T}$ to be ``material" or ``radiative" determines whether one calls the phase transition ``purely ferroelectric" or ``superradiant", and this in turn is determined by the gauge choice as discussed earlier. 

We finally consider a concrete example. We assume that each dipole has canonical operators pointing along ${\bm \varepsilon}$ 
and a double-well potential $V(\theta,\phi) = -\theta r^2/2 + \phi r^4/4$ where $\theta$ and $\phi$ control the shape of the double-well. The Hamiltonian of each dipole is therefore 
$H_m^\alpha = {{\cal E}\over 2}\left(-\partial_\zeta^2 -\beta\zeta^2+{\zeta^4 \over 2} \right)$ 
\cite{de_bernardis_breakdown_2018} where we have defined $\zeta = r/r_0$ with $r_0=(1/[m\phi])^{1/6}$, along with ${\cal E}=1/(mr_0^2)$ and $\beta=\theta m r_0^4$. We also define the gauge-invariant dimensionless coupling parameter $\eta = (e/\omega)\sqrt{\rho/m}$. The parameters $e,\,m$ and $\rho$ can now be eliminated in favour of ${\cal E}, \beta$ and $\eta$.
%
%
\begin{figure}[t]
\centering
%
\hspace*{1mm}
\includegraphics[scale=0.417]{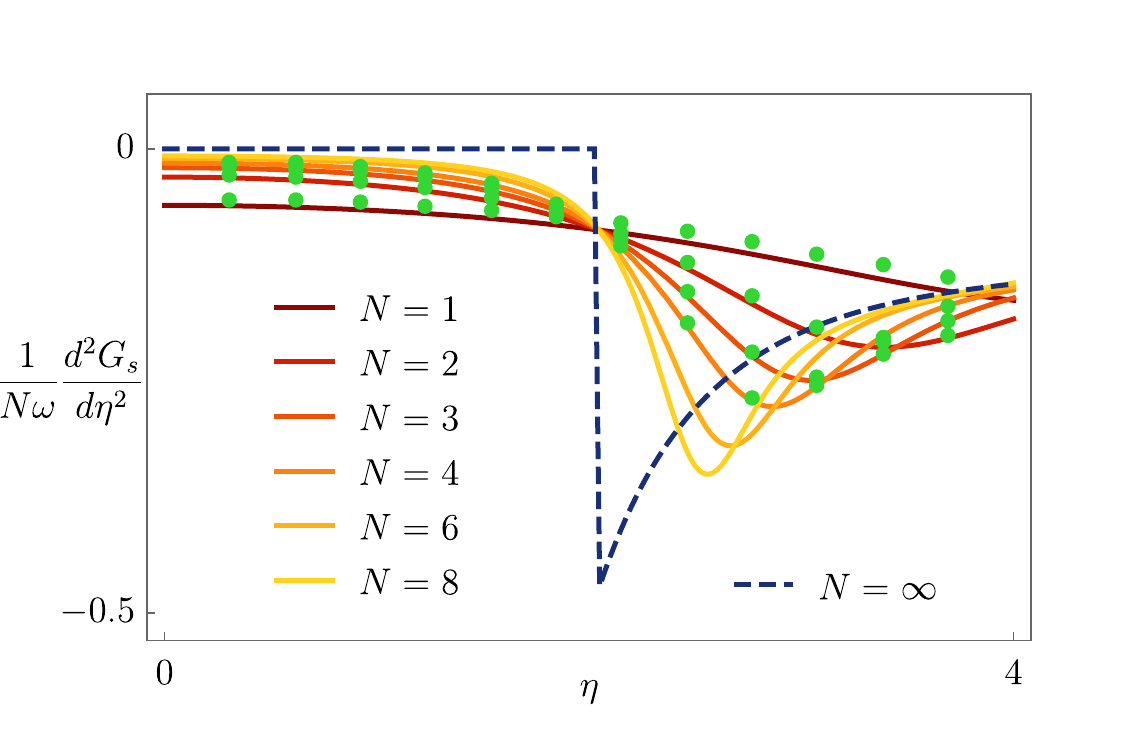}
\vspace*{-0.3cm}
\caption{
Second derivative of the normalised ground energy 
$(N\omega)^{-1}d^2 G_s/d\eta^2$ plotted for various values of $N$ as a function of $\eta$ found using the multipolar-gauge two-level model (solid curves). A precursor to the discontinuity that locates the phase transition in the limit 
$N=\infty$ can clearly be seen. The green dotted curves provide exact (gauge-invariant) predictions found without two-level truncation. 
Agreement already occurs at $N=4$. $\beta=3.3$ and ${\cal E}$ is chosen such that $\omega_m=\omega$.}
\label{trans}
\vspace*{-0.3cm}
\end{figure}
%
To demonstrate that it is possible to obtain accurate two-level model predictions and to show a clear precursor to the phase transition, in Fig.~\ref{trans} we consider the normalised second derivative of the (shifted) ground energy \cite{emary_quantum_2003}. In the abnormal phase of the thermodynamic limit this is given by
\begin{align}
{1\over N\omega} {d^2 G_s^{\alpha,2}\over d\eta^2}\bigg|_{\rm th,a} = -\frac{\omega_m}{\omega}{d^2\over d\eta^2}{(1-\tau)^2\over 4\tau}
\end{align}
where $G_s=G-\rho d^2/2$ is the ground energy $G$ shifted by the coupling-dependent term $\rho d^2/2$ in Eq. (\ref{h2}). 
We choose $\beta=3.3$, which provides a highly anharmonic single-dipole spectrum such that $(\epsilon_2-\epsilon_0)/\omega_m \approx 36$. The two-level truncation within the multipolar gauge $\alpha=1$ is subsequently found to be accurate in predicting low energy properties. This was first confirmed in the case of the Rabi model $N=1$ in Ref.~\cite{de_bernardis_breakdown_2018}. The accuracy actually increases with $N$ and convergence of exact (gauge-invariant, no two-level truncation) and approximate predictions already occurs at $N=4$. 
The situation may change if the double-well is parameterised differently such that the multipolar truncation is not optimal \cite{stokes_gauge_2019}, 
see also additional analysis in~\cite{Note1}. 

The situation may also change if additional cavity modes are taken into account \cite{munoz_resolution_2018,roth_optimal_2019}. In particular, the multipolar-gauge coupling scales as $\sqrt{\omega}$ such that the single-mode approximation appears least favourable in this gauge, and has been shown to breakdown in the ultrastrong-coupling regime \cite{munoz_resolution_2018}. To incorporate some of the effects of non-resonant modes within a Dicke-type model 
a formal procedure of adiabatic elimination 
can be used and this also has the advantage of enabling an exploration of more diverse dipolar geometries \cite{de_bernardis_cavity_2018}. 
Nevertheless, for our purpose of determining whether a physical phase transition can be supported by systems describable using a Dicke model and on understanding 
its relationship to the choice of gauge, the single-mode restriction 
is sufficient, because the qualitative behaviour of the thermodynamic limit of the single-mode Dicke model is known to carry over to the multi-mode case \cite{hepp_equilibrium_1973,pimentel_phase_1975}. The extension to general dipolar arrangements, and to more sophisticated cavity models warrants further study.


We have shown that a unique physical phase transition can occur in simple many-dipole cavity QED systems. 
We have resolved all ambiguities pertaining to the choice of gauge by determining both the origin and properties of the phase transition in terms of any gauge's definitions of the quantum subsystems, and by demonstrating equivalence between all gauge choices. We have shown that the original ``no-go theorem"~\cite{rzazewski_phase_1975} does not apply, and also that one need not look beyond ordinary cavity QED in order to find systems supporting a superradiant phase transition. A no-go {\em theorem} for ground state superradiance occurs for, and only for, the Coulomb-gauge definition of radiation. We have shown that although the two-level approximation ruins the gauge-invariance of the theory, unambiguous predictions can still be obtained. 
The framework developed here should be straightforwardly extendable to artificial solid-state and superconducting systems, as well as to driven and dissipative systems. 
This will elucidate both qualitatively and quantitatively the underlying causes and physical natures of thermodynamic phase transitions therein, and in each case, 
determine optimal approximate descriptions.
%
%
%
\begin{acknowledgments}
This work was supported by the UK Engineering and Physical Sciences Research Council, grant no. EP/N008154/1. We thank J. Keeling and P. Rabl for useful correspondence, 
and Z. Blunden-Codd, M. Mitchison, R. Puebla, and D. De Bernardis for useful discussions.
\end{acknowledgments}


%

\onecolumngrid

\newpage

\begin{center}
\textbf{Supplemental Material - Uniqueness of the phase transition in many-dipole cavity QED systems}\\
\vspace*{3mm}
Adam Stokes and Ahsan Nazir
\end{center}

\section*{
Arbitrary gauge quantisation of the matter-radiation system}\label{gi}

Throughout this section we will frequently use the Helmholtz decomposition of a vector field ${\bf V}$ into transverse and longitudinal parts ${\bf V}_{\rm T}$ and ${\bf V}_{\rm L}$ such that for all ${\bf x}$
\begin{align}
&{\bf V}={\bf V}_{\rm T}+{\bf V}_{\rm L},\\ &\nabla \cdot {\bf V}_{\rm T}({\bf x})=0,\\ &\nabla \times {\bf V}_{\rm L}({\bf x}) ={\bf 0}.
\end{align}
We assume that all vector fields vanish at the boundaries $|{\bf x}|\to \infty$, which allows free use of integration by parts such as
\begin{align}\label{intbpts}
\int d^3 x\, {\bf V}({\bf x})\cdot \nabla f({\bf x}) = -\int d^3 x \, f({\bf x})\nabla \cdot {\bf V}({\bf x}).
\end{align}
Recalling that $\nabla \times \nabla f({\bf x}) ={\bf 0}$ for any $f$ and for all ${\bf x}$, we have that for any longitudinal field ${\bf U}_{\rm L}$ there exists an $f$ such that  ${\bf U}_{\rm L}=\nabla f$. It follows from Eq. (\ref{intbpts}) that
\begin{align}\label{vu}
 \int d^3 x\, {\bf V}_{\rm T}({\bf x})\cdot {\bf U}_{\rm L}({\bf x}) =0
\end{align}
for any vector fields ${\bf V}$ and ${\bf U}$. These formulae will be frequently used in what follows.

We consider $N$ hydrogen-like non-relativistic atoms comprised of positive charges $e_{\rm p}=e$ with mass $m_{\rm p}$ at positions ${\bf r}_{\rm p\mu}$, each of which is paired with a negative charge $e_{\rm e}=-e$ with mass $m_{\rm e}$ at ${\bf r}_{\rm e \mu}$. The four-current has components $(j^a)=(\rho, {\bf J})$ with
\begin{align}
\rho({\bf x}) &= \sum_{\mu=1}^N \rho_\mu({\bf x}) = \sum_{\mu=1}^N\sum_{\sigma={\rm e,p}}e_\sigma\delta({\bf x}-{\bf r}_{\rm \sigma\mu})= e\sum_{\mu=1}^N \left[\delta({\bf x}-{\bf r}_{\rm p\mu})-\delta({\bf x}-{\bf r}_{\rm e\mu})\right], \\
{\bf J}({\bf x}) &= \sum_{\mu=1}^N {\bf J}_\mu({\bf x}) = \sum_{\mu=1}^N\sum_{\sigma={\rm e,p}}e_\sigma{\dot {\bf r}}_{\sigma\mu}\delta({\bf x}-{\bf r}_{\rm \sigma\mu})=e\sum_{\mu =1}^N \left[{\dot {\bf r}}_{\rm p\mu} \delta({\bf x}-{\bf r}_{\rm p\mu})-{\dot {\bf r}}_{\rm e\mu} \delta({\bf x}-{\bf r}_{\rm e\mu})\right],
\end{align}
such that $\partial_a j^a = {\dot \rho}({\bf x})+\nabla \cdot {\bf J}=0$. Since the system is globally neutral we can define the polarisation field ${\bf P}$ by the equation $-\nabla \cdot {\bf P} = \rho$, which can be solved to give
\begin{align}\label{pol}
{\bf P}({\bf x}) = -\int d^3 x'\, {\bf g}({\bf x},{\bf x}')\rho({\bf x}') = -\sum_{\mu=1}^N \int d^3 x'\, {\bf g}({\bf x},{\bf x}')\rho_\mu({\bf x}') \equiv \sum_{\mu=1}^N  {\bf P}_\mu({\bf x})
\end{align} 
where
\begin{align}\label{g}
\nabla \cdot {\bf g}({\bf x},{\bf x}') = \delta ({\bf x}-{\bf x}')
\end{align}
defines the green's function ${\bf g}$ for the divergence operator. Since $\nabla \cdot {\bf g}({\bf x},{\bf x'}) \equiv \nabla \cdot {\bf g}_{\rm L}({\bf x},{\bf x}')$, Eq. (\ref{pol}) only fixes ${\bf g}_{\rm L}={\bf g}-{\bf g}_{\rm T}$ uniquely as
\begin{align}\label{gl}
{\bf g}_{\rm L}({\bf x},{\bf x}') = -\nabla {1\over 4\uppi |{\bf x}-{\bf x}'|}.
\end{align}
Any field ${\bf g}_{\rm T}$ with $\nabla \cdot {\bf g}_{\rm T}({\bf x},{\bf x}')=0$, can be added to ${\bf g}_{\rm L}$ in Eq. (\ref{gl}) to obtain a ${\bf g}$ that satisfies Eq. (\ref{pol}). It follows that ${\bf P}_{\rm L}$ is fixed uniquely by Eqs. (\ref{pol}) and (\ref{gl}) while ${\bf P}_{\rm T}$ is arbitrary, being determined by Eq. (\ref{pol}) and ${\bf g}_{\rm T}$. To avoid any confusion we note that here we are using the notation ${\bf P}_{\rm T}$ to refer to the arbitrary transverse part of ${\bf P}$ given in Eq. (\ref{pol}). In the main text we reserve the notation ${\bf P}_{\rm T}$ for specifically the {\em multipolar} transverse polarisation, which is just one possible example of the transverse part of ${\bf P}$.

Assuming for generality an external potential $V_{\rm ext}$ acting on the charges the standard Lagrangian describing the system of all charges coupled to the Maxwell field is given by
\begin{align}\label{lag1}
L = {1\over 2}\sum_{\mu =1}^N \left[m_{\rm p} {\dot {\bf r}}_{\rm p\mu}^2+m_{\rm e}{\dot {\bf r}}_{\rm e\mu}^2\right] -V_{\rm ext} -\int d^3 x\, \left[ j^a({\bf x}) A_a({\bf x})+{1\over 4}F_{ab}({\bf x})F^{ab}({\bf x})\right]
\end{align}
where $(A^a)=(A_0,{\bf A})$ are the components of the electromagnetic four-potential and $F_{ab} = \partial_a A_b - \partial_b A_a$. The field tensor components $F_{ab}$ are invariant under a gauge transformation $A_a \to A_a - \partial_a \chi$ where $\chi$ is an arbitrary function. We encode this gauge-freedom into the arbitrary function ${\bf g}_{\rm T}$ by defining
\begin{align}\label{chi}
\chi({\bf x}) = \int d^3 x'\, {\bf g}({\bf x}',{\bf x})\cdot {\bf A}_{\rm T}({\bf x}') \equiv \int d^3 x'\, {\bf g}_{\rm T}({\bf x}',{\bf x})\cdot {\bf A}_{\rm T}({\bf x}')
\end{align} 
and by subsequently defining the arbitrary potentials
\begin{align}
A_0 &= \phi_{\rm Coul} -\partial_t \chi,\label{a0} \\
{\bf A} &= {\bf A}_{\rm T} + {\bf A}_{\rm L} \equiv {\bf A}_{\rm T} + \nabla \chi \label{avec}
\end{align}
where ${\bf A}_{\rm T}$ is the gauge-invariant transverse vector potential and 
\begin{align}
\phi_{\rm Coul}({\bf x}) = \int d^3 x'\, {\rho({\bf x'})\over 4\pi |{\bf x}-{\bf x}'|}.
\end{align}
The choice ${\bf g}_{\rm T} ={\bf 0}$ defines the Coulomb gauge wherein ${\bf A}={\bf A}_{\rm T}$ and $A_0 =\phi_{\rm Coul}$. A different gauge can by specified by choosing a different ${\bf g}_{\rm T}$.

The Lagrangian $L$ in Eq.~(\ref{lag1}) is not gauge-invariant. We therefore define the equivalent, but gauge-invariant Lagrangian
\begin{align}\label{lag2}
L '= L - {d\over dt}\int d^3 x\, {\bf P}({\bf x})\cdot {\bf A}({\bf x}).
\end{align}
Using Eqs.~(\ref{a0}) and (\ref{avec}) in conjunction with Eq.~(\ref{chi}) the Lagrangian $L'$ can be written
\begin{align}\label{Lp}
L'=L_0 - {d\over dt}\int d^3 x\, {\bf P}_{\rm T}({\bf x})\cdot {\bf A}_{\rm T}({\bf x})
\end{align}
where
\begin{align}\label{L0}
L_0 =  {1\over 2}\sum_{\mu =1}^N \left[m_{\rm p} {\dot {\bf r}}_{\rm p\mu}^2+m_{\rm e}{\dot {\bf r}}_{\rm e\mu}^2\right] -V_{\rm ext} -{1\over 2}\int d^3 x\, \rho({\bf x})\phi_{\rm Coul}({\bf x})+\int d^3 x\, {\bf J}({\bf x}) \cdot {\bf A}_{\rm T}({\bf x})+{1\over 2}\int d^3 x\, \left[{\bf E}_{\rm T}({\bf x})^2-{\bf B}({\bf x})^2\right]
\end{align}
in which ${\bf E}_{\rm T}=-{\dot {\bf A}}_{\rm T}$ and ${\bf B}=\nabla \times {\bf A}_{\rm T}$. The remaining total time derivative in Eq. (\ref{Lp}) depends on ${\bf P}_{\rm T}$ which according to Eq. (\ref{pol}) is uniquely determined through a choice of gauge ${\bf g}_{\rm T}$. It is straightforward to show that
\begin{align}\label{pchi}
\int d^3 x\, {\bf P}_{\rm T}({\bf x})\cdot {\bf A}_{\rm T}({\bf x})= - \int d^3 x\, \rho({\bf x})\chi({\bf x}),
\end{align}
where now according to Eq. (\ref{chi}) it is the arbitrary function $\chi$ that is determined by the gauge ${\bf g}_{\rm T}$. Note that ${\bf P}_{\rm T}$ in Eq.~(\ref{Lp}) is a completely arbitrary transverse field and need not coincide with the usual multipolar transverse polarisation field. In writing Eq.~(\ref{L0}) we have used the Gauss law $\nabla\cdot {\bf E}_{\rm L}= -\nabla^2\phi_{\rm Coul}= \rho$ and integration by parts to separate the transverse and electrostatic parts of the electromagnetic Lagrangian as
\begin{align}\label{es}
-{1\over 4}\int d^3 x \, F_{ab}F^{ab}= {1\over 2} \int d^3 x\, \left[{\bf E}({\bf x})^2-{\bf B}({\bf x})^2\right] = {1\over 2} \int d^3 x\, \left[{\bf E}_{\rm T}({\bf x})^2-{\bf B}({\bf x})^2\right] +{1\over 2} \int d^3 x\, \rho({\bf x})\phi_{\rm Coul}({\bf x}).
\end{align}
Combining the electrostatic part of Eq~(\ref{es}) with the term $-\int d^3 x \rho\phi_{\rm Coul}$ coming from the component $-\int d^3 x A_0 \rho$ of the interaction Lagrangian we obtain the final electrostatic interaction term $-\int d^3 x \rho\phi_{\rm Coul}/2$ that appears in Eq.~(\ref{L0}).

We now introduce relative and centre-of-mass coordinates for each atom $\mu$, which are defined by
\begin{align}
{\bf r}_\mu &= {\bf r}_{\rm e\mu}-{\bf r}_{\rm p\mu},\\
{\bf R}_\mu &= {m_{\rm e}{\bf r}_{\rm e\mu}+m_{\rm p}{\bf r}_{\rm p\mu} \over m_{\rm e}+m_{\rm p}},
\end{align}
along with total and reduced masses defined by
\begin{align}
M&=m_{\rm e}+m_{\rm p},\\
m&={m_{\rm e}m_{\rm p}\over M}.
\end{align}
We partition the electrostatic interaction term into intra-atomic and inter-atomic contributions as
\begin{align}\label{coul}
{1\over 2}\int d^3 x\, \rho({\bf x})\phi_{\rm Coul}({\bf x}) = \int d^3 x \int d^3 x' {\rho({\bf x})\rho({\bf x}') \over 8\pi |{\bf x}-{\bf x}'|} = \sum_{\mu=1}^N V_\mu + \sum_{\mu \neq \nu}^N \int d^3 x \int d^3 x'{\rho_\mu ({\bf x})\rho_\nu ({\bf x}') \over 8\pi |{\bf x}-{\bf x}'|}
\end{align}
where
\begin{align}\label{sel}
V_\mu = V_\mu^{\rm self} - \sum_{\mu=1}^N {e^2 \over 4\pi |{\bf r}_\mu|}.
\end{align}
The first term on the right-hand-side in Eq. (\ref{sel}) gives the divergent Coulomb self-energies of the charges at ${\bf r}_{\rm e\mu}$ and ${\bf r}_{\rm p\mu}$. The second term gives the potential energy that binds the charge $-e$ at ${\bf r}_{\rm e\mu}$ to its nucleus $+e$ at ${\bf r}_{\rm p\mu}$. The second term on the right-hand-side of Eq.~(\ref{coul}) gives the inter-atomic Coulomb interactions.

We now restrict our attention to gauges in which the arbitrary transverse polarisation takes the form of a weighted multipolar transverse polarisation with weight $\alpha$;
\begin{align}\label{PT}
P_{{\rm T},i}({\bf x}) = \sum_{\mu =1}^N P_{{\rm T}\mu,i}({\bf x}) = \alpha \sum_{\mu =1}^N \sum_{\sigma={\rm e,p}}e_\sigma ({\bf r}_{\sigma\mu} -{\bf R}_\mu)_j \int_0^1 d\lambda \,\delta_{ij}^{\rm T}({\bf x}-{\bf R}_\mu-\lambda [{\bf r}_{\sigma \mu}-{\bf R}_\mu])  =: P_{{\rm T}\alpha,i}({\bf x})
\end{align}
where $\alpha$ is real and dimensionless and where ${\bf P}_{{\rm T}\mu}$ denotes the transverse polarisation associated with the $\mu$'th atom expressed in terms of the centre-of-mass position ${\bf R}_\mu$. With this restriction the gauge is now completely determined by selecting a value of $\alpha$. The derivation of the Dicke-model also requires the electric-dipole approximation (EDA), which for simplicity we implement at this stage rather than later on. In the EDA the charge density within the inter-atomic Coulomb interaction is approximated by the first non-zero term in the multipole expansion of the single atom density $\rho_\mu$ about the atomic centre-of-mass at ${\bf R}_\mu$, viz.,
\begin{align}\label{multrho}
\rho({\bf x}) &=\sum_{\mu =1}^N \rho_\mu({\bf x}) = \sum_{\mu =1}^N \sum_{\sigma={\rm e,p}}e_\sigma \delta({\bf x}-{\bf R}_\mu-({\bf r}_{\sigma\mu}-{\bf R}_\mu))\nonumber \\ &= \sum_{\mu =1}^N \sum_{\sigma={\rm e,p}}e_\sigma \left[\delta({\bf x}-{\bf R}_\mu)-({\bf r}_{\sigma \mu}-{\bf R}_\mu)\cdot \nabla\delta({\bf x}-{\bf R}_\mu)+...\right] \approx -\sum_{\mu =1}^N {\bf d}_\mu \cdot \nabla \delta({\bf x}-{\bf R}_\mu)
\end{align}
where ${\bf d}_\mu = -e{\bf r}_\mu$ is the dipole moment of the $\mu$'th dipole. We thereby obtain
\begin{align}\label{dd}
\sum_{\mu \neq \nu}^N \int d^3x \int d^3 x'{\rho_\mu ({\bf x})\rho_\nu ({\bf x}') \over 8\pi |{\bf x}-{\bf x}'|} \approx {1\over 2} \sum_{\mu \neq \nu}^N d_{\mu,i}d_{\nu,j}\delta_{ij}^{\rm L}({\bf R}_\mu-{\bf R}_\nu) =: V_{\rm dip}.
\end{align}
Similarly to Eq. (\ref{multrho}) the multipole expansion of the current yields to leading order
\begin{align}
{\bf J}({\bf x}) &=\sum_{\mu =1}^N {\bf J}_\mu({\bf x}) = \sum_{\mu =1}^N \sum_{\sigma={\rm e,p}}e_\sigma {\dot {\bf r}}_{\sigma\mu}\delta({\bf x}-{\bf R}_\mu-({\bf r}_{\sigma\mu}-{\bf R}_\mu))\nonumber \\ &= \sum_{\mu =1}^N \sum_{\sigma={\rm e,p}}e_\sigma  {\dot {\bf r}}_{\sigma\mu}\left[\delta({\bf x}-{\bf R}_\mu)-({\bf r}_{\sigma \mu}-{\bf R}_\mu)\cdot \nabla\delta({\bf x}-{\bf R}_\mu)+...\right] \approx \sum_{\mu =1}^N {\dot {\bf d}}_\mu \delta({\bf x}-{\bf R}_\mu).
\end{align}
The current-dependent interaction component of $L_0$ in Eq. (\ref{L0}) therefore becomes in the EDA
\begin{align}\label{curreda}
\int d^3 x \, {\bf J}({\bf x})\cdot {\bf A}_{\rm T}({\bf x}) \approx \sum_{\mu =1}^N {\dot {\bf d}}_\mu \cdot {\bf A}_{\rm T}({\bf R}_\mu).
\end{align}
The multipole expansion of the $\alpha$-dependent transverse polarisation field in Eq. (\ref{PT}) yields to leading order
\begin{align}\label{PTeda}
P_{{\rm T}\alpha,i}({\bf x}) &= \alpha \sum_{\mu =1}^N \sum_{\sigma={\rm e,p}}e_\sigma  ({\bf r}_{\sigma\mu} -{\bf R}_\mu)\int_0^1 d\lambda \left[1-\lambda ({\bf r}_{\sigma\mu}-{\bf R}_\mu)\cdot \nabla +...\right]\delta_{ij}^{\rm T}({\bf x}-{\bf R}_\mu) \nonumber \\ &\approx \alpha \sum_{\mu=1}^N d_{\mu,j}\delta_{ij}^{\rm T}({\bf x} - {\bf R}_\mu).
\end{align}
At this stage we neglect the nuclear motions ${\dot {\bf R}}_\mu$, such that when we arrive at the quantum theory ${\bf R}_\mu$ will simply denote the fixed classical position of the $\mu$'th dipole. The polarisation-dependent interaction component of $L'$ in Eq. (\ref{Lp}) therefore becomes
\begin{align}
-{d\over dt}\int d^3 x\, {\bf P}_{\rm T\alpha}({\bf x})\cdot {\bf A}_{\rm T}({\bf x}) =-\alpha {d\over dt} \sum_{\mu=1}^N {\bf d}_\mu\cdot {\bf A}_{\rm T}({\bf R}_\mu) =-\alpha \sum_{\mu=1}^N \left[{\dot {\bf d}}_\mu\cdot {\bf A}_{\rm T}({\bf R}_\mu) +{\bf d}_\mu\cdot {\dot {\bf A}}_{\rm T}({\bf R}_\mu) \right].
\end{align}
Altogether Eqs. (\ref{dd}), (\ref{curreda}) and (\ref{PTeda}) yield the Lagrangian $L'$ within the EDA as
\begin{align}\label{lageda}
L' = \sum_{\mu =1}^N {1\over 2}m {\dot {\bf r}}_\mu^2 -V -V_{\rm dip} + \sum_{\mu=1}^N \left[(1-\alpha){\dot {\bf d}}_\mu \cdot {\bf A}_{\rm T}({\bf R}_\mu) - \alpha {\bf d}_\mu \cdot {\dot {\bf A}}_{\rm T}({\bf R}_\mu)\right]+{1\over 2}\int d^3 x\, \left[{\bf E}_{\rm T}({\bf x})^2-{\bf B}({\bf x})^2\right] =: L_\alpha
\end{align}
where we have absorbed the external potential $V_{\rm ext}$ into the definition of the total intra-atomic potential as
\begin{align}
V := \sum_{\mu=1}^N V_\mu +V_{\rm ext}.
\end{align}
The electrostatic energies $V_\mu$ and $V_{\rm dip}$ are defined in Eqs.~(\ref{sel}) and (\ref{dd}) respectively, while the transverse electric and magnetic fields are given by ${\bf E}_{\rm T}=-{\dot {\bf A}}_{\rm T}$ and ${\bf B}=\nabla\times {\bf A}_{\rm T}$ respectively. Thus, the Lagrangian in Eq. (\ref{lageda}) is fully specified in terms of the dynamical variable set $\{{\bf r}_\mu,~{\dot {\bf r}}_\mu, {\bf A}_{\rm T},~{\dot {\bf A}}_{\rm T}\}$ together with the fixed dipolar positions ${\bf R}_\mu$.

It is now possible to switch to the canonical formalism by defining the canonical momenta
\begin{align}
{\bf p}_{\mu\alpha} &= {\partial L_\alpha \over \partial {\dot {\bf r}}_\mu}, \\
{\bf \Pi}_{\rm T\alpha} &= {\delta L_\alpha \over \delta {\dot {\bf A}}_{\rm T}}
\end{align}
and to then quantise the theory by assuming the canonical commutation relations
\begin{align}
[r_{\mu,i},p_{\nu,j}]&=i\delta_{\mu\nu}\delta_{ij} \label{com1},\\
[A_{{\rm T},i}({\bf x}),\Pi_{{\rm T},j}({\bf x}')]&=i\delta_{ij}^{\rm T}({\bf x}-{\bf x}').\label{com2}
\end{align}
The centre-of-mass variable ${\bf R}_\mu$ is a classical position of the $\mu$'th dipole. The $\alpha$-dependent canonical momenta are found to be
\begin{align}
{\bf p}_{\mu\alpha} &= m{\dot {\bf r}}_\mu - e(1-\alpha){\bf A}_{\rm T}({\bf R}_\mu),\label{mom1}\\
{\bf \Pi}_{{\rm T}\alpha} &= -{\bf E}_{\rm T} - {\bf P}_{{\rm T}\alpha}.\label{mom2}
\end{align}
For any two values $\alpha$ and $\alpha'$ of the gauge parameter the canonical operators are related by the unitary gauge-fixing transformation $R_{\alpha\alpha'}$ as
\begin{align}
{\bf p}_{\mu\alpha} &= R_{\alpha\alpha'}{\bf p}_{\mu\alpha'} R_{\alpha\alpha'}^{-1},\\
{\bf \Pi}_{{\rm T}\alpha} &= R_{\alpha\alpha'}{\bf \Pi}_{{\rm T}\alpha'} R_{\alpha\alpha'}^{-1}
\end{align}
where
\begin{align}\label{gt}
R_{\alpha\alpha'} = \exp\left[i(\alpha-\alpha')\sum_{\mu=1}^N {\bf d}_\mu \cdot {\bf A}_{\rm T}({\bf R}_\mu)\right].
\end{align}

The Hamiltonian is defined by
\begin{align}\label{hamdef}
H = \sum_{\mu=1}^N {\dot {\bf r}}_\mu \cdot {\bf p}_{\mu\alpha} + \int d^3 x \, {\dot {\bf A}}_{\rm T}({\bf x})\cdot {\bf \Pi}_{{\rm T}\alpha}({\bf x})-L_\alpha.
\end{align}
Through substitution of Eqs.~(\ref{mom1}) and (\ref{mom2}) into Eq. (\ref{hamdef}) the Hamiltonian written in terms of the manifestly gauge-invariant operators $\{{\bf r}_\mu, {\dot {\bf r}}_\mu,{\bf A}_{\rm T},{\dot {\bf A}}_{\rm T}\}$ is found to coincide with the total energy expressed as the sum of material and transverse-electromagnetic energies;
\begin{align}
H = E_{\rm matter}+E_{\rm field} \equiv \left[ \sum_{\mu =1}^N  {1\over 2}m {\dot {\bf r}}_\mu^2 +V +V_{\rm dip} \right] + \left[{1\over 2}\int d^3 x\, \left[{\bf E}_{\rm T}({\bf x})^2+{\bf B}({\bf x})^2\right]\right].
\end{align}
While this expression is clearly $\alpha$-independent (gauge-invariant), when expressed in terms of canonical operators the Hamiltonian has an $\alpha$-dependent functional form given by
\begin{align}\label{ham1}
H = &\sum_{\mu=1}^N {1\over 2m}\left[{\bf p}_{\mu\alpha}+e(1-\alpha){\bf A}_{\rm T}({\bf R}_\mu)\right]^2 + V +{\alpha^2\over 2}\sum_{\mu=1}^N d_{\mu,i}d_{\mu,j}\delta_{ij}^{\rm T}({\bf 0}) + (1-\alpha^2)V_{\rm dip} +\alpha \sum_{\mu=1}^N {\bf d}_\mu \cdot {\bf \Pi}_{{\rm T}\alpha}({\bf R}_\mu)\nonumber  \\ &+ {1\over 2}\int d^3 x\, \left[{\bf \Pi}_{{\rm T}\alpha}({\bf x})^2+{\bf B}({\bf x})^2\right].
\end{align}
In writing Eq. (\ref{ham1}) we have used
\begin{align}
{1\over 2}\int d^3 x {\bf P}_{{\rm T}\alpha}({\bf x})^2 = {\alpha^2\over 2} \sum_{\mu,\nu=1}^N d_{\mu,i}d_{\nu,j}\delta_{ij}^{\rm T}({\bf R}_\mu-{\bf R}_\nu) = {\alpha^2\over 2} \sum_{\mu=1}^N d_{\mu,i}d_{\nu,j}\delta_{ij}^{\rm T}({\bf 0}) -\alpha^2 V_{\rm dip},
\end{align}
which follows from Eqs. (\ref{dd}) and (\ref{PTeda}), together with the property $\delta^{\rm T}_{ij}({\bf x})=-\delta^{\rm L}_{ij}({\bf x})$ for ${\bf x}\neq {\bf 0}$. The Hamiltonian reduces to the Coulomb-gauge result if we choose $\alpha=0$. In this case direct electrostatic interactions are fully explicit in the form of the dipole-dipole interaction $(1-\alpha^2)V_{\rm dip}=V_{\rm dip}$. The field degrees of freedom are defined in terms of the transverse vector potential and its velocity, the transverse electric field; ${\bf \Pi}_{{\rm T}0}={\dot {\bf A}}_{\rm T}=-{\bf E}_{\rm T}$. Another common choice of gauge is the multipolar gauge obtained by choosing $\alpha=1$. In this gauge electrostatic interactions are eliminated; $(1-\alpha^2)V_{\rm dip}=0$, while the field degrees of freedom are defined in terms of the transverse vector potential and the retarded transverse displacement field; ${\bf \Pi}_{{\rm T}1}=-{\bf D}_{\rm T}= -{\bf E}_{\rm T}-{\bf P}_{{\rm T}1}$. Outside of the atoms the transverse displacement field coincides with the total electric field, which in the EDA means that ${\bf D}_{\rm T}({\bf x})= {\bf E}({\bf x})$ for  ${\bf x}\neq {\bf R}_\mu$. More generally, in the $\alpha$-gauge the Hamiltonian has a hybrid form. Coulomb-gauge matter-transverse field interaction terms are weighted by $(1-\alpha)$ while multipolar-gauge matter-transverse field interaction terms are weighted by $\alpha$. Electrostatic interaction terms are weighted by $1-\alpha^2$. In the case $N=2$ the Hamiltonian in Eq.~(\ref{ham1}) coincides with that given in Ref. \cite{stokes_master_2018}.

The above expressions are applicable for general field operators ${\bf A}_{\rm T}$ and ${\bf \Pi}_{{\rm T}\alpha}$. We now define the operator
\begin{align}
a_{\alpha,\lambda}({\bf k}) := \sqrt{1\over 2\omega}\bigg(\omega {\tilde A}_{{\rm T},\lambda}({\bf k})+{\rm i}{\tilde \Pi}_{{\rm T}\alpha,\lambda}({\bf k})\bigg)
\end{align}
where ${\tilde A}_{{\rm T},\lambda}({\bf k})={\bm \varepsilon}_\lambda({\bf k})\cdot {\tilde {\bf A}}_{{\rm T}}({\bf k})$ and ${\tilde \Pi}_{{\rm T}\alpha,\lambda}({\bf k})={\bm \varepsilon}_\lambda({\bf k})\cdot {\tilde {\bf \Pi}}_{{\rm T}\alpha}({\bf k})$. Here tildes denote the Fourier transform and ${\bm \varepsilon}_{\lambda}({\bf k}),~\lambda=1,\,2$ are mutually orthogonal unit vectors both orthogonal to ${\bf k}$.
From the transverse canonical commutation relation
\begin{align}\label{tcom}
[A_{{\rm T},i}({\bf x}),\Pi_{{\rm T}\alpha,j}({\bf x}')]={\rm i}\delta_{ij}^{\rm T}({\bf x}-{\bf x}')
\end{align}
it follows that
\begin{align}\label{boscom2}
[a_{\alpha,\lambda}({\bf k}),a_{\alpha,\lambda'}^\dagger({\bf k}')]=\delta_{\lambda\lambda'}\delta({\bf k}-{\bf k}').
\end{align}
The operators $a_{\alpha,\lambda}({\bf k})$ and $a_{\alpha,\lambda}^\dagger({\bf k})$ are recognisable as annihilation and creation operators for a photon with momentum ${\bf k}$ and polarisation $\lambda$. In terms of these operators the canonical fields support the Fourier representations
\begin{align}\label{modeex}
&{\bf A}_{\rm T}({\bf x}) = \int d^3 k \sum_\lambda g{\bm \varepsilon}_\lambda({\bf k})\left(a^\dagger_{\alpha,\lambda}({\bf k}){\rm e}^{-{\rm i}{\bf k}\cdot {\bf x}} + a_{\alpha,\lambda}({\bf k}){\rm e}^{{\rm i}{\bf k}\cdot {\bf x}}\right),\nonumber \\ &
{\bm \Pi}_{{\rm T}\alpha}({\bf x}) = {\rm i}\int d^3 k \sum_\lambda \omega g{\bm \varepsilon}_\lambda({\bf k})\left(a^\dagger_{\alpha,\lambda}({\bf k}){\rm e}^{-{\rm i}{\bf k}\cdot {\bf x}} - a_{\alpha,\lambda}({\bf k}){\rm e}^{i{\bf k}\cdot {\bf x}}\right)
\end{align}
where $\omega=|{\bf k}|$ and $g:={1 / \sqrt{2\omega (2\uppi)^3}}$.

If we assume an implicit cavity with volume $v$ that satisfies periodic boundary conditions, the continuous label ${\bf k}$ becomes discrete. The pair ${\bf k}\lambda$ then labels a radiation mode and the operators $a_{\alpha, \lambda}({\bf k})$ are labelled with discrete index $\cdot_{\bf k}$ as $a_{\alpha, {\bf k}\lambda}$, which satisfy
\begin{align}\label{boscom3}
[a_{\alpha,{\bf k}\lambda},a_{\alpha,{\bf k}'\lambda'}^\dagger]=\delta_{\lambda\lambda'}\delta_{{\bf k}{\bf k}'}.
\end{align}
As a less realistic, but simpler model for the cavity we may restrict our attention to a single fixed mode ${\bf k}\lambda$ and ignore all modes ${\bf k}'\lambda'\neq {\bf k}\lambda$. In this case the field operators become
\begin{align}
&{\bf A}_{\rm T}({\bf x}) =  g {\bm \varepsilon} \left(a_\alpha^\dagger {\rm e}^{-{\rm i}{\bf k}\cdot {\bf x}}+a_\alpha {\rm e}^{{\rm i}{\bf k}\cdot {\bf x}}\right),\label{consm}\\ &{\bf \Pi}_{{\rm T}\alpha}({\bf x}) ={\rm i} \omega g {\bm \varepsilon} \left(a_\alpha^\dagger {\rm e}^{-{\rm i}{\bf k}\cdot {\bf x}}-a_\alpha {\rm e}^{{\rm i}{\bf k}\cdot {\bf x}}\right),\label{consm2}
\end{align}
where $g=1/\sqrt{2\omega v}$, $\omega=|{\bf k}|$, ${\bm \varepsilon} \equiv {\bm \varepsilon}_{{\bf k}\lambda}$. and $a_\alpha \equiv a_{\alpha,{\bf k}\lambda}$ with $[a_\alpha,a_\alpha^\dagger]=1$. Eqs. (\ref{consm}) and (\ref{consm2}) imply that the cavity canonical operators now satisfy the commutation relation 
\begin{align}\label{tcom2}
[A_{{\rm T},i}({\bf x}),\Pi_{{\rm T}\alpha,j}({\bf x}')]={{\rm i}\varepsilon_i \varepsilon_j \over v}\cos \left[{\bf k}\cdot ({\bf x}-{\bf x}')\right].
\end{align}
To preserve Eq. (\ref{tcom}) we must discretise the ${\bf k}$-space representation of the transverse delta-function and subsequently perform the single-mode approximation as
\begin{align}\label{del2}
\delta^{\rm T}_{ij}({\bf x}) =  \int {d^3 k\over (2\uppi)^3} \sum_{\lambda=1,2} \varepsilon_{\lambda,i}({\bf k}) \varepsilon_{\lambda,j}({\bf k}) \cos({\bf k}\cdot {\bf x}) \longrightarrow  \sum_{{\bf k}\lambda} {\varepsilon_{{\bf k}\lambda,i} \varepsilon_{{\bf k}\lambda,j} \over v} \cos({\bf k}\cdot {\bf x}) \longrightarrow {\varepsilon_i \varepsilon_j \over v} \cos({\bf k}\cdot {\bf x}).
\end{align}
With this the $\alpha$-gauge transverse material polarisation becomes
\begin{align}\label{pol1}
{\bf P}_{{\rm T}\alpha}({\bf x}) = {\alpha \over v} \sum_{\mu=1}^N {\bm \varepsilon} ({\hat {\bf d}}_\mu \cdot {\bm \varepsilon}) \cos\left[{\bf k}\cdot ({\bf x}-{\bf R}_\mu)\right].
\end{align}
Similarly, $V_{\rm dip}$ becomes
\begin{align}\label{vdip1}
V_{\rm dip} =- {1\over 2} \sum_{\mu \neq \nu}^N d_{\mu,i}d_{\nu,j}\delta_{ij}^{\rm T}({\bf R}_\mu-{\bf R}_\nu) =-{1\over 2v} \sum_{\mu\neq \nu}^N ({\bf d_\mu}\cdot {\bm \varepsilon})({\bf d_\nu}\cdot {\bm \varepsilon})\cos\left[{\bf k}\cdot ({\bf R}_\mu -{\bf R}_\nu)\right].
\end{align}
Altogether, within the single-mode theory the Hamiltonian in Eq. (\ref{ham1}) becomes
\begin{align}\label{ham2}
H= &\sum_{\mu=1}^N {1\over 2m}\left[{\bf p}_{\mu\alpha}+e(1-\alpha){\bf A}_{\rm T}({\bf R}_\mu)\right]^2 + V +{\alpha^2\over 2v}\sum_{\mu=1}^N ({\bf d_\mu}\cdot {\bm \varepsilon})^2 + (1-\alpha^2)V_{\rm dip} +\alpha \sum_{\mu=1}^N {\hat {\bf d}}_\mu \cdot {\bf \Pi}_{{\rm T}\alpha}({\bf R}_\mu) +\omega\left(a_\alpha^\dagger a_\alpha +{1\over 2} \right).
\end{align}
where ${\bf A}_{\rm T}({\bf R}_\mu)$, ${\bf \Pi}_{{\rm T}\alpha}({\bf R}_\mu)$ and $V_{\rm dip}$ are given by Eqs.~(\ref{consm}), (\ref{consm2}) and (\ref{vdip1}) respectively.

Within the single-mode restriction the Heisenberg equation with the Hamiltonian in Eq. (\ref{ham2}) yields
\begin{align}
{\bf E}_{\rm T}({\bf x})=-{\dot {\bf A}}_{\rm T}({\bf x})=-{\rm i} \omega g {\bm \varepsilon} \left(a_\alpha^\dagger  {\rm e}^{-{\rm i}{\bf k}\cdot {\bf x}}-a_\alpha  {\rm e}^{{\rm i}{\bf k}\cdot {\bf x}}\right)- {\alpha \over v} \sum_{\mu=1}^N {\bm \varepsilon}({\bf d}_\mu \cdot {\bm \varepsilon})\cos\left[{\bf k}\cdot ({\bf x}-{\bf R}_\mu)\right]=-{\bf \Pi}_{{\rm T}\alpha}({\bf x})-{\bf P}_{{\rm T}\alpha}({\bf x})
\end{align}
as required according to Eq.~(\ref{mom2}). Because the single-mode restriction has been imposed on both the mode operators and the material transverse polarisation the fundamental kinematic relations of the Hamiltonian theory, namely Eqs. (\ref{mom1}) and (\ref{mom2}), are preserved. We therefore obtain a self-consistent theory describing $N$ dipoles and a single-mode of radiation. We note that as in the multi-mode theory of Eq. (\ref{ham1}) electrostatic inter-dipole interactions are explicit in Eq.~(\ref{ham2}) in all gauges other than the multipolar gauge $\alpha=1$. 

Finally we consider the limit of closely spaced dipoles around the origin ${\bf 0}$ such that ${\bf R}_\mu \approx {\bf 0}$. In this case, according to Eqs. (\ref{pol1}) and (\ref{vdip1}) we obtain
\begin{align}
V_{\rm dip} &= -{1\over 2v} \sum_{\mu\neq \nu}^N ({\hat {\bf d}}_\mu \cdot {\bm \varepsilon})({\hat {\bf d}}_\nu \cdot {\bm \varepsilon}),\label{vdipfin}\\
{\bf P}_{{\rm T}\alpha}&:={\bf P}_{{\rm T}\alpha}({\bf 0})={\alpha \over v} \sum_{\mu=1}^N {\bm \varepsilon}({\hat {\bf d}}_\mu \cdot {\bm \varepsilon})
\end{align}
The canonical fields ${\bf A}_{\rm T}({\bf R}_\mu)$ and ${\bf \Pi}({\bf R}_\mu)$ within the Hamiltonian are replaced by ${\bf A}:={\bf A}_{\rm T}({\bf 0})$ and ${\bf \Pi}_\alpha:={\bf \Pi}_{\rm T\alpha}({\bf 0})$ respectively, where for notational convenience we have dropped the transversality subscript $\cdot_{\rm T}$. We also adopt this convention in the main text. 

\section*{
Diagonalisation of generic bilinear coupled-oscillator Hamiltonian, normal-phase, and abnormal-phase Hamiltonians}

Here we diagonalise a coupled oscillator Hamiltonian with the generic structure that we will repeatedly encounter. We define arbitrary oscillator operators $\{y,\,y^\dagger,\,z,\,z^\dagger\}$ with $[y,y^\dagger]=1=[z,z^\dagger]$ and where all other commutators between elements of $\{y,\,y^\dagger,\,z,\,z^\dagger\}$ vanish. The generic Hamiltonian we wish to diagonalise is
\begin{align}\label{genh}
h:=\, &w y^\dagger y + w'z^\dagger z + ig(y^\dagger + y)(z'^\dagger -z') -i g'(y^\dagger-y)(z^\dagger +z) +C
\end{align}
where $C$ is a constant. 
As an example, the Hamiltonian in Eq.~(\ref{hth}), which describes the thermodynamic limit of the $\alpha$-gauge Dicke model in the normal phase, has the form of $H$ above.

To diagonalise $h$ we introduce Hermitian quadratures $q_\mu = (\mu^\dagger+\mu)/\sqrt{2}$ and $p_\mu=i(\mu^\dagger-\mu)/\sqrt{2}$ where $\mu=y,\, z$. Subsequently we define the tuple of quadratures ${\bf r}=(q_y,q_z,p_y,p_z)$, which is such that $[r_j,r_k]=i\Omega_{jk}$ where
\begin{align}
\Omega = \left( {\begin{array}{cc}
   0 & I \\
   -I & 0 \\
  \end{array} } \right)
\end{align}
is a matrix representation of the standard symplectic form on ${\mathbb R}^{2\times 2}$. The Hamiltonian in Eq.~(\ref{genh}) can now be written
\begin{align}
h= {\bf r}^{\rm T}M{\bf r} +C-{1\over 2}(w+w')
\end{align}
where $\cdot^{\rm T}$ denotes transposition and
\begin{align}\label{M}
 M=
 {1\over 2} \left( {\begin{array}{cccc}
    w & 0 & 0 & 2g \\
   0 & w' & -2g' & 0 \\
   0 & -2g' & w & 0 \\
   2g & 0 & 0 & w' \\
  \end{array} } \right)
\end{align}
is assumed to be positive-definite. By Williamson's theorem \cite{williamson_algebraic_1936} there exists a symplectic matrix $\Lambda$ such that
\begin{align}
\Lambda^{\rm T} M \Lambda ={\tilde D}= \left( {\begin{array}{cc}
   D & 0 \\
   0 & D \\
  \end{array} } \right)
\end{align}
where $D$ is diagonal. Denoting the elements of $D$ by $\nu_j,\,j=1,2$, the quantity $\pm i\nu_j$ is an eigenvalue of $\Omega M$. We therefore make use of the canonically transformed quadratures ${\bf r}'=(q'_y,q'_z,p'_y,p'_z)=\Lambda^{-1}{\bf r}$, which because $\Lambda$ is symplectic also satisfy $[r'_j,r'_k]=i\Omega_{jk}$. The Hamiltonian $h$ can now be written in terms of upper and lower polaritons as
\begin{align}
h= {\bf r'}^{\rm T}{\tilde D}{\bf r'} +C-{1\over 2}(w+w')= E_+ y'^\dagger y' + E_- z'^\dagger z' + {1\over 2}( E_+ +E_- - w-w') + C
\end{align}
where $y',\,z'$ are bosonic operators defined in terms of the transformed quadratures ${\bf r}'$. They satisfy $[y',y'^\dagger]=1=[z',z'^\dagger]$ while all other commutators between elements of $\{y',\,y'^\dagger,\,z',\,z'^\dagger\}$ vanish. The energies $E_\pm$ are given by the elements of $D$ as $E_+^2=(2\nu_1)^2$ and $E_-^2=(2\nu_2)^2$. The $\nu_j$ are found from the matrix $\Omega M$, which is found from Eq.~(\ref{M}). Explicitly, the polariton energies $E_\pm$ are given by
\begin{align}
2E_\pm^2 = 8gg'+w^2+w'^2\pm\sqrt{(w^2-w'^2)^2+16(wg'+w'g)(wg+w'g')}.
\end{align}


The $\alpha$-gauge Dicke model Hamiltonian is
\begin{align}\label{hd}
H^{\alpha,2} =&\, \omega_m J_\alpha^z +{N\over 2}(\epsilon_0+\epsilon_1) +{1\over 2}\rho d^2+\omega_\alpha \left(c_\alpha^\dagger c_\alpha+{1\over 2}\right) -{{\cal C}_\alpha \over N}\left(J_\alpha^++J_\alpha^-\right)^2\nonumber \\ &-i(1-\alpha)\omega_m d _\alpha\sqrt{\rho \over 2\omega_\alpha N} (J_\alpha^+-J_\alpha^-)(c_\alpha^\dagger +c_\alpha) +i\alpha d \sqrt{\rho \omega_\alpha \over 2 N} (J_\alpha^++J_\alpha^-)(c_\alpha^\dagger -c_\alpha).
\end{align}
where to obtain this expression the transverse vector potential and its conjugate momentum at the dipolar positions ${\bf R}_\mu$ have been approximated by their values at the origin as described in the previous section, and they are given by
\begin{align}
{\bf A}_{\rm T}({\bf 0}) &={{\bm \varepsilon}\over \sqrt{2\omega v}}(a_\alpha^\dagger +a_\alpha) =  {{\bm \varepsilon}\over \sqrt{2\omega_\alpha v}}(c_\alpha^\dagger +c_\alpha),\\
{\bf \Pi}_\alpha({\bf 0}) &=i{\bm \varepsilon}\sqrt{\omega \over 2 v}(a_\alpha^\dagger -a_\alpha) = i{\bm \varepsilon}\sqrt{\omega_\alpha \over 2 v}(c_\alpha^\dagger -c_\alpha).
\end{align}
The bosonic operators $c_\alpha$ include the contribution of the ${\bf A}^2$-term of the Hamiltonian implicitly. This is why the frequency appearing in the corresponding mode expansions above is the renormalised frequency $\omega_\alpha$ and why there is no explicit ${\bf A}^2$-term within the Hamiltonian in Eq.~(\ref{hd}). The $c_\alpha$ are related to the unrenormalised bosonic operators $\alpha_\alpha$ by a local Bogoliubov tranformation within the cavity Hilbert space.

Substituting
\begin{align}
&J_\alpha^z = b_\alpha^\dagger b_\alpha -{N\over 2} \nonumber \\
&J_\alpha^+ = b_\alpha^\dagger \sqrt{N-b_\alpha^\dagger b_\alpha},~~~J_\alpha^-=(J_\alpha^+)^\dagger
\end{align} 
into Eq.~(\ref{hd}) gives
\begin{align}\label{hint}
H^{\alpha,2} =&\, \omega_m \left(b_\alpha^\dagger b_\alpha -{N\over 2}\right) +{N\over 2}(\epsilon_0+\epsilon_1) +{1\over 2}\rho d^2+\omega_\alpha \left(c_\alpha^\dagger c_\alpha+{1\over 2}\right)  -{\cal C}_\alpha \left(b_\alpha^\dagger \sqrt{1-{b_\alpha^\dagger b_\alpha\over N}}+\sqrt{1-{b_\alpha^\dagger b_\alpha\over N}}b_\alpha\right)^2\nonumber \\ &-i(1-\alpha)\omega_m d _\alpha\sqrt{\rho \over 2\omega_\alpha } \left(b_\alpha^\dagger \sqrt{1-{b_\alpha^\dagger b_\alpha\over N}}-\sqrt{1-{b_\alpha^\dagger b_\alpha\over N}}b_\alpha\right)(c_\alpha^\dagger +c_\alpha) \nonumber \\ &+i\alpha d \sqrt{\rho \omega_\alpha \over 2} \left(b_\alpha^\dagger \sqrt{1-{b_\alpha^\dagger b_\alpha\over N}}+\sqrt{1-{b_\alpha^\dagger b_\alpha\over N}}b_\alpha\right)(c_\alpha^\dagger -c_\alpha).
\end{align}
All terms that depend on the square-root functions of the mode operators $b_\alpha,\,b_\alpha^\dagger$ have coefficients that remain finite in the thermodynamic limit. Therefore, expanding the square-roots as 
\begin{align}
\sqrt{1-{b_\alpha^\dagger b_\alpha \over N}}=1-{b_\alpha^\dagger b_\alpha \over 2N}+...
\end{align}
and ignoring terms which vanish in the thermodynamic limit ($N\to \infty$) constitutes making the replacement
\begin{align}
\sqrt{1-{b_\alpha^\dagger b_\alpha \over N}}\to 1
\end{align}
in Eq.~(\ref{hint}), which yields
\begin{align}\label{hth}
H_{\rm th}^{\alpha,2} =&\, \omega_m \left(b_\alpha^\dagger b_\alpha -{N\over 2}\right) +{N\over 2}(\epsilon_0+\epsilon_1) +{1\over 2}\rho d^2+\omega_\alpha \left(c_\alpha^\dagger c_\alpha+{1\over 2}\right)  -{\cal C}_\alpha \left(b_\alpha^\dagger + b_\alpha\right)^2\nonumber \\ &-i(1-\alpha)\omega_m d _\alpha\sqrt{\rho \over 2\omega_\alpha } \left(b_\alpha^\dagger -b_\alpha\right)(c_\alpha^\dagger +c_\alpha) +i\alpha d \sqrt{\rho \omega_\alpha \over 2} \left(b_\alpha^\dagger +b_\alpha\right)(c_\alpha^\dagger -c_\alpha).
\end{align}
We now combine the terms that depend only on $b_\alpha,~b_\alpha^\dagger$ via a Bogoliubov transformation such that
\begin{align}
\omega_m b_\alpha^\dagger b_\alpha - {\cal C}_\alpha \left[b_\alpha^\dagger + b_\alpha\right]^2 = {\tilde \omega}_m^\alpha l_\alpha^\dagger l_\alpha + {1\over 2}\left({\tilde \omega}_m^\alpha-\omega_m\right)
\end{align}
where $[l_\alpha,l_\alpha^\dagger]=1$ and
\begin{align}
{{\tilde \omega}_m^\alpha}^2 = \omega_m^2 - 4{\cal C}_\alpha.
\end{align}
The quadrature operators defined in terms of the $b_\alpha$ are given in terms of the $l_\alpha$ by
\begin{align}
b_\alpha^\dagger -b_\alpha &= \sqrt{{\tilde \omega}_m^\alpha \over \omega_m}(l_\alpha^\dagger -l_\alpha),\label{l1} \\ 
b_\alpha^\dagger +b_\alpha &= \sqrt{\omega_m\over {\tilde \omega}_m^\alpha}(l_\alpha^\dagger +l_\alpha).\label{l2}
\end{align}
Substituting these expressions into Eq.~(\ref{hth}) gives
\begin{align}\label{hth2}
H_{\rm th}^{\alpha,2} =\, {\tilde \omega}_m^\alpha l_\alpha^\dagger l_\alpha +\omega_\alpha c_\alpha^\dagger c_\alpha + C_\alpha -i(1-\alpha) d _\alpha\sqrt{\omega_m {\tilde \omega}_m^\alpha \rho \over 2\omega_\alpha } \left(l_\alpha^\dagger -l_\alpha\right)(c_\alpha^\dagger +c_\alpha) +i\alpha d \sqrt{\rho \omega_\alpha \omega_m  \over 2  {\tilde \omega}_m^\alpha} \left(l_\alpha^\dagger +l_\alpha\right)(c_\alpha^\dagger -c_\alpha)
\end{align}
where we have combined all constant terms into
\begin{align}
C_\alpha = N\epsilon_0 +{1\over 2}\left({\tilde \omega}_m^\alpha-\omega_m+\omega_\alpha +\rho d^2\right).
\end{align}
Note that if we choose the material potential such that $\epsilon_0=0$, i.e., if the material energy zero-point is zero, then $C_\alpha$ is $N$-independent and remains finite in the limit $N\to \infty$. The Hamiltonian in Eq.~(\ref{hth2}) can be diagonalised using the method presented at the start of this section, which leads to the final result given in the main text.


To find the abnormal-phase Hamiltonian, the relevant starting Hamiltonian $H^{\alpha,2}$ is again that of Eq. (\ref{hd}), together with the Holstein-Primakoff representation
\begin{align}
&J_\alpha^z = b_\alpha^\dagger b_\alpha -{N\over 2} \nonumber \\
&J_\alpha^+ = b_\alpha^\dagger \sqrt{N-b_\alpha^\dagger b_\alpha},~~~J_\alpha^-=(J_\alpha^+)^\dagger.
\end{align} 
We then introduce the displaced operator $f_\alpha = b_\alpha-\sqrt{\beta_\alpha}$ where $\beta_\alpha$ is assumed to be of order $N$, such that the material part $H_m^{\alpha,2}$ of $H^{\alpha,2}$ can be written
\begin{align}\label{hm0}
H_{m}^{\alpha,2}&=\omega_m b_\alpha^\dagger b_\alpha - {{\cal C}_\alpha\over N}\left(b_\alpha^\dagger \sqrt{N-b_\alpha^\dagger b_\alpha}+\sqrt{N-b_\alpha^\dagger b_\alpha}b_\alpha\right)^2 \nonumber \\ &= \omega_m\left(f_\alpha^\dagger f_\alpha -\sqrt{\beta_\alpha}[f_\alpha^\dagger +f_\alpha]+\beta_\alpha\right) - {\cal C}_\alpha{N-\beta_\alpha\over N} \left([f_\alpha^\dagger-\sqrt{\beta_\alpha}] \sqrt{\xi_\alpha}+\sqrt{\xi_\alpha}[f_\alpha-\sqrt{\beta_\alpha}]\right)^2
\end{align}
where for convenience we have defined
\begin{align}\label{xi}
\xi_\alpha:= 1-{f_\alpha^\dagger f_\alpha -\sqrt{\beta_\alpha}[f_\alpha^\dagger +f_\alpha]\over N-\beta_\alpha}.
\end{align}
Expanding $\sqrt{\xi_\alpha}$ in Eq. (\ref{hm0}) and neglecting terms which vanish in the thermodynamic limit one obtains after lengthy manipulations
\begin{align}\label{h2thm0}
H_{{\rm th},m}^{\alpha,2} =& \left( \omega_m + 4{\cal C}_\alpha{\beta_\alpha \over N}\right)f_\alpha^\dagger f_\alpha -{{\cal C}_\alpha\over N}(N-5\beta_\alpha)\left(f_\alpha^\dagger +f_\alpha\right)^2 + \beta_\alpha \omega_m - 2{\cal C}_\alpha\left({N-\beta_\alpha\over N}\right)\left(2\beta_\alpha +{\beta_\alpha\over N-\beta_\alpha}\right) \nonumber\\ &-\sqrt{\beta_\alpha}\left(\omega_m -4{\cal C}_\alpha{N-2\beta_\alpha \over N}\right)\left(f_\alpha^\dagger +f_\alpha\right).
\end{align}
Next we replace the radiation mode operators with displaced operators such that $c_\alpha = i(k_\alpha+\sqrt{\gamma_\alpha})$. The total Hamiltonian therefore reads
\begin{align}
H_{\rm th}^{\alpha,2} = & H_{{\rm th},m}^{\alpha,2} + \omega_\alpha \left(k_\alpha^\dagger k_\alpha +\sqrt{\gamma_\alpha}[k^\dagger_\alpha+k_\alpha]+\gamma_\alpha+{1\over 2}\right)-g_\alpha'\sqrt{N-\beta_\alpha\over N}(f_\alpha^\dagger\sqrt{\xi_\alpha}-\sqrt{\xi_\alpha}f_\alpha)(k_\alpha^\dagger-k_\alpha) \nonumber \\ &+g_\alpha\sqrt{N-\beta_\alpha \over N}\left(f_\alpha^\dagger \sqrt{\xi_\alpha}+\sqrt{\xi_\alpha}  f_\alpha -2\sqrt{\beta_\alpha\xi_\alpha}\right)\left(k_\alpha^\dagger +k_\alpha +2\sqrt{\gamma_\alpha}\right) + {1\over 2}\rho d^2+N\epsilon_0
\end{align}
We now collect all terms that are linear in the $f_\alpha$ or in the $k_\alpha$ and choose $\beta_\alpha$ and $\gamma_\alpha$ such that these terms vanish. The trivial case in which $\beta_\alpha=0=\gamma_\alpha$ yields the normal phase Hamiltonian. We will see that the non-trivial solutions
\begin{align}
\beta_\alpha&=\beta := {N\over 2}(1-\tau),\label{bets}\\
\gamma_\alpha &= {N g_\alpha^2 \over \omega_\alpha^2}\left(1-\tau^2\right)\label{gams}
\end{align}
where
\begin{align}
\tau:= {\omega_\alpha \omega_m \over 4(g_\alpha^2+\omega_\alpha {\cal C}_\alpha)}= {\alpha^2 \omega_\alpha \omega_m \over 4g_\alpha^2} = {\omega_m \over 2\rho d^2}.
\end{align}
yield a Hamiltonian describing the abnormal phase.

Expanding $\sqrt{\xi_\alpha}$ and neglecting terms which vanish in the thermodynamic limit one now obtains after lengthy manipulations
\begin{align}
H_{\rm th}^{\alpha,2} =& \left( \omega_m + 4{\cal C}_\alpha{\beta \over N}+2g_\alpha \sqrt{\gamma_\alpha \beta \over N(N-\beta)}\right)f_\alpha^\dagger f_\alpha +\omega_\alpha k_\alpha^\dagger k_\alpha \nonumber \\ &+\left(\sqrt{\gamma_\alpha \beta \over N(N-\beta)}\left[1+{\beta\over 2(N-\beta)}\right]-{{\cal C}_\alpha\over N}(N-5\beta)\right)(f_\alpha^\dagger +f_\alpha)^2-g_\alpha'\sqrt{N-\beta\over N}(f_\alpha^\dagger -f_\alpha)(k_\alpha^\dagger-k_\alpha) \nonumber \\ &+g_\alpha\sqrt{N-\beta \over N}\left(1-{\beta\over N-\beta}\right)\left(f_\alpha^\dagger +  f_\alpha \right)\left(k_\alpha^\dagger +k_\alpha\right) \nonumber \\ & +N\epsilon_0 +{1\over 2}\rho d^2 + {\omega_\alpha\over 2}+\gamma_\alpha\omega_\alpha + \beta \omega_m -2{{\cal C}_\alpha \beta \over N}(1+2[N-\beta]) -g_\alpha \sqrt{\gamma_\alpha \beta \over N(N-\beta)}\left(1+4(N-\beta)\right)
\end{align}
We can remove the term quadratic in $f_\alpha^\dagger+f_\alpha$ by defining new material mode operators $f_\alpha',\, f_\alpha'^\dagger$ such that
\begin{align}
&\left( \omega_m + 4{\cal C}_\alpha{\beta \over N}+2g_\alpha \sqrt{\gamma_\alpha \beta \over N(N-\beta)}\right)\left(f_\alpha^\dagger f_\alpha+{1\over 2}\right) \nonumber \\ &+\left(\sqrt{\gamma_\alpha \beta \over N(N-\beta)}\left[1+{\beta\over 2(N-\beta)}\right]-{{\cal C}_\alpha\over N}(N-5\beta)\right)(f_\alpha^\dagger +f_\alpha)^2 \nonumber \\ =&\, {\omega_m\over 2\tau}(1+\tau)\left(f_\alpha^\dagger f_\alpha+{1\over 2}\right)+\left({{\cal C}_\alpha \over 2}[3-5\tau]+{\omega_m \alpha^2(1-\tau)(3+\tau) \over 8\tau(1+\tau)}\right)(f_\alpha^\dagger +f_\alpha)^2 \nonumber \\ =&\, \underaccent{\tilde}{\omega}_m^\alpha\left(f_\alpha'^\dagger f_\alpha'+{1\over 2}\right)
\end{align}
where
\begin{align}
{\underaccent{\tilde}{\omega}_m^\alpha}^2 &= {{\omega_m}^2\over \tau^2}\left[1-(1-\alpha^2)\tau^2\right].
\end{align}
Letting $\omega_m'=\omega_m(1+\tau)/(2\tau)$ and using the relations
\begin{align}
&f_\alpha^\dagger-f_\alpha = \sqrt{\underaccent{\tilde}{\omega}_m^\alpha\over \omega_m'}(f_\alpha'^\dagger-f_\alpha')\nonumber \\ 
&f_\alpha^\dagger+f_\alpha = \sqrt{\omega_m'\over \underaccent{\tilde}{\omega}_m^\alpha}(f_\alpha'^\dagger+f_\alpha') \nonumber \\ 
\end{align}
the Hamiltonian can be written
\begin{align}\label{abhs}
H_{\rm th}^{\alpha,2} = \underaccent{\tilde}{\omega}_m^\alpha f_\alpha'^\dagger f_\alpha' +\omega_\alpha {c'}_\alpha^\dagger c'_\alpha - i\underaccent{\tilde}{g}'_\alpha(f_\alpha'^\dagger -f_\alpha')({c'}_\alpha^\dagger +c'_\alpha)+i\underaccent{\tilde}{g}_\alpha(f_\alpha'^\dagger +f'_\alpha)({c'}_\alpha^\dagger -c'_\alpha) + C'_\alpha
\end{align}
where $c'_\alpha=i k_\alpha$ while
\begin{align}
\underaccent{\tilde}{g}'_\alpha &= \sqrt{\tau\underaccent{\tilde}{\omega}_m^\alpha \over \omega_m}g_\alpha',\\
\underaccent{\tilde}{g}_\alpha &= \sqrt{\tau \omega_m \over \underaccent{\tilde}{\omega}_m^\alpha}g_\alpha,
\end{align}
and
\begin{align}
C_\alpha'=N\left(\epsilon_0-{g_\alpha^2\over \omega_\alpha \alpha^2}[1-\tau^2]+{\omega_m \over 2}[1-\tau]\right)-{g_\alpha^2\over \alpha^2 \omega_\alpha}(1-\tau)+{\omega_\alpha\over 2} +{\underaccent{\tilde}{\omega}_m^\alpha \over 2} -{\omega_m'\over 2}+{1\over 2}\rho d^2.
\end{align}
The Hamiltonian in Eq. (\ref{abhs}) has the form in Eq. (\ref{genh}) and can therefore be diagonalised by the method presented at the start of this section, leading to the final result denoted $H_{2,{\rm th}}^{\alpha,{\rm a}}$, and given in 
the main text.

\section*{
Polariton Energies}

\begin{figure}[H]
  \centering
\hspace*{-3mm}(a)\includegraphics[scale=0.393]{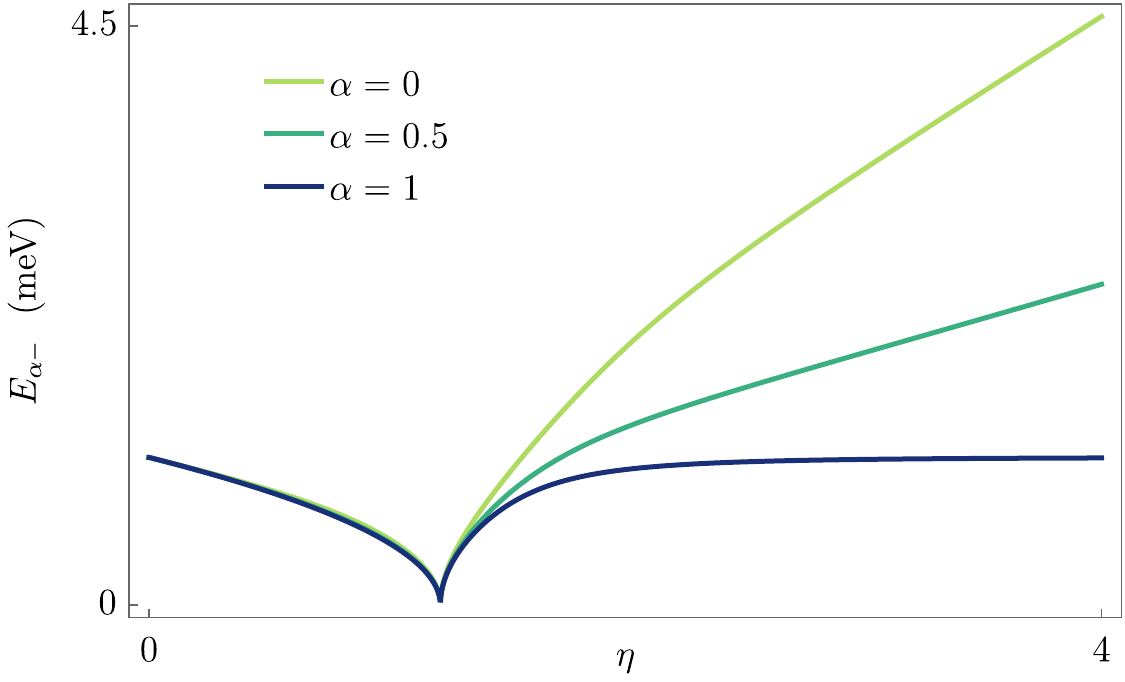}
  \vspace*{0.3cm}

\hspace*{-4mm}(b)\hspace*{1.5mm}\includegraphics[scale=0.383]{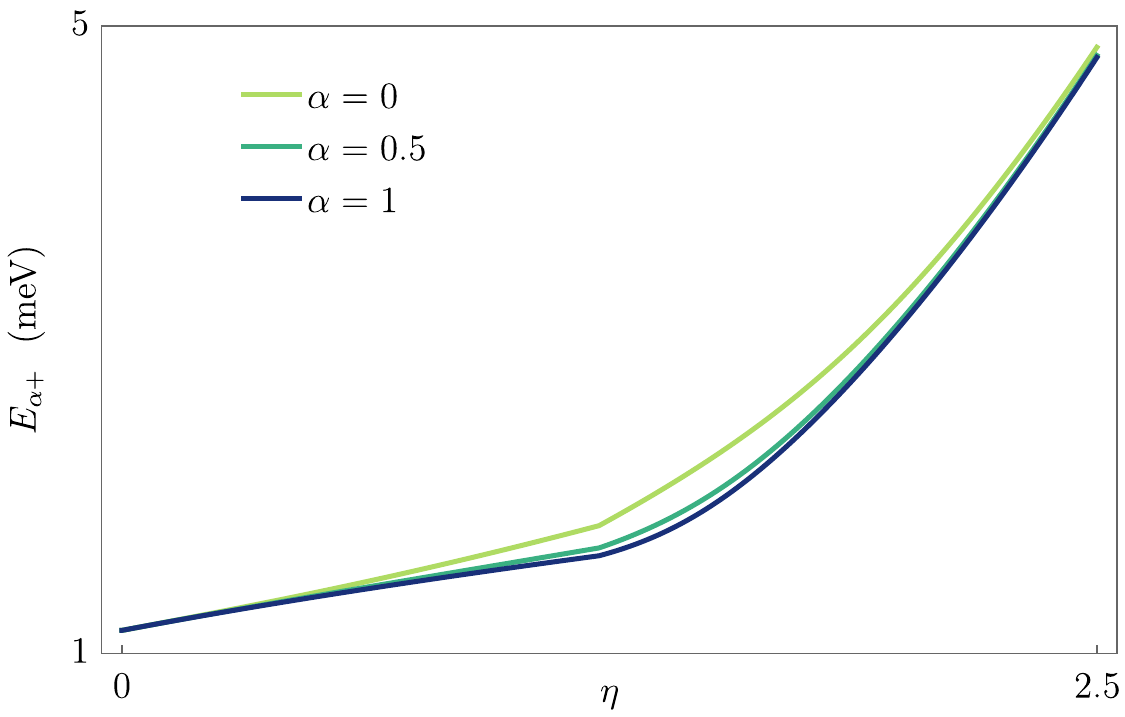}
  \caption{In all plots we have chosen $\beta=2.4$, and then chosen ${\cal E}$ such that $\omega_m=\omega$. (a) The lower polariton energy is plotted for three values of $\alpha$ as a function of $\eta$. The qualitative behaviour of $E_{\alpha-}$ as $\eta$ becomes large depends on the value of $\alpha$. (b) The upper polariton energy is plotted for three values of $\alpha$ as a function of $\eta$. As with the lower polariton energy, due to the two-level truncation the behaviour depends on the value of $\alpha$.}
  \label{energies}
\end{figure}

Here we plot the polariton energies $E_\alpha^\pm$ in Fig.~\ref{energies} using the example of double-well dipoles as considered in the main text. The thermodynamic limit of the Dicke-model is seen to be gauge-dependent, i.e., the $E_\alpha^\pm$ are $\alpha$-dependent. This occurs due to the use of material level truncation. Despite this, it is clearly seen that the occurrence of a unique phase transition is obtained as a gauge-invariant prediction. Furthermore, we are able to determine its gauge-invariant manifestation, which is plotted in the following section.

\section*{
Calculation of radiative canonical operator averages}

\begin{figure}[H]
  \centering
\hspace*{-2mm}\includegraphics[scale=0.393]{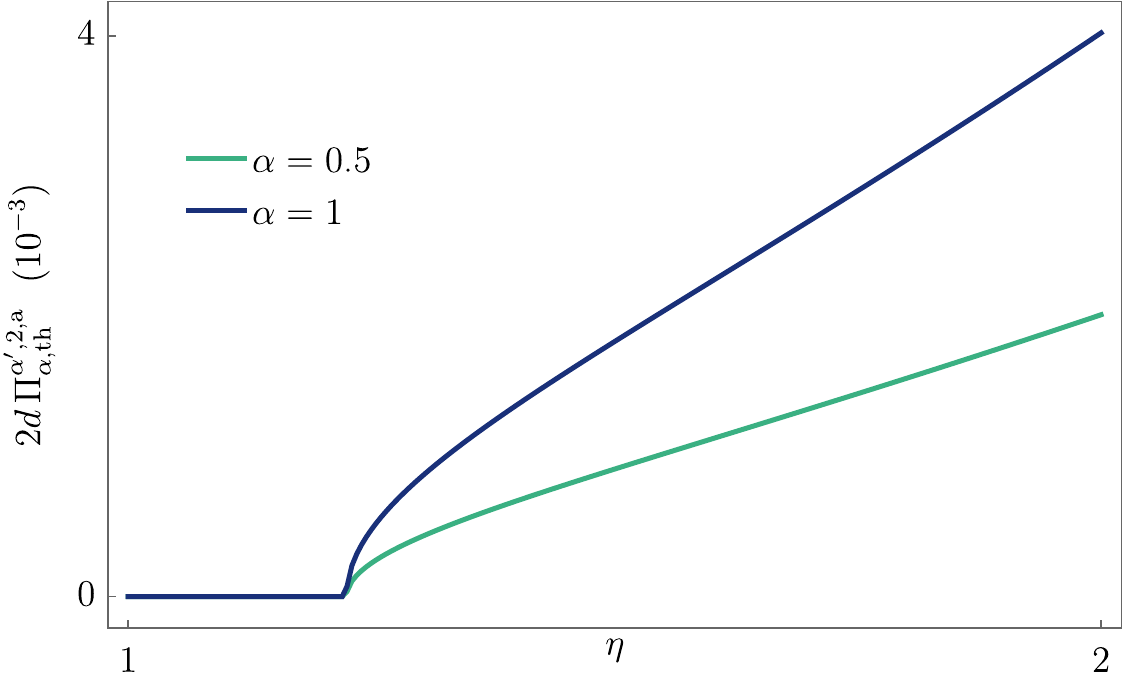}
  \vspace*{0.3cm}
  \caption{The quantity $2d\,\Pi_{\alpha,{\rm th}}^{\alpha',2}$ is plotted as a function of $\eta$ for $\alpha = \alpha_{\rm JC},~1$. As in Fig. \ref{energies}, we have chosen $\beta=2.4$, and then chosen ${\cal E}$ such that $\omega_m=\omega$. 
The definition of the canonical momentum changes linearly with $\alpha$ from $\alpha=0$ such that ${\bf \Pi}_0=-{\bf E}_{\rm T}$ to $\alpha=1$ such that ${\bf \Pi}_1=-{\bf D}_{\rm T}$. Correspondingly, for fixed $\alpha'$ we have $\Pi_{\alpha,{\rm th}}^{\alpha',2}/\Pi_{\alpha'',{\rm th}}^{\alpha',2}=\alpha/\alpha''$. Thus, the ratio of the magnitudes of the two curves is always $\alpha_{\rm JC}/1=\alpha_{\rm JC}$. Note in addition, that $\Pi_{1,{\rm th}}^{\alpha',2}= P_{\rm T,th}^{\alpha',2}$, meaning that the curve corresponding to $\alpha=1$ illustrates the gauge-invariant manifestation of the abnormal phase via the transverse polarisation ${\bf P}_{\rm T}$.}
  \label{momf}
\end{figure}

We begin with the cavity canonical operators $A={\bm\varepsilon}\cdot {\bf A}$ and $\Pi_\alpha={\bm\varepsilon}\cdot {\bf \Pi}_\alpha$. The ground state average of $A$ is trivially zero. Similarly the ground state average of $\Pi_\alpha$ in the normal phase is zero. The ground state average of $\Pi_\alpha$ in the abnormal phase can be calculated using the Dicke-model of any gauge $\alpha'$. We begin with the expression
\begin{align}\label{Y2s}
\Pi_\alpha^{\alpha',2} = \Pi_{\alpha'}-{d\over v}(\alpha-\alpha')(J_{\alpha'}^+ + J_{\alpha'}^-),
\end{align}
which is the $\alpha'$-gauge's two-level approximation of $\Pi_\alpha$. Using the Holstein-Primakoff representation and then defining the displaced operators $f_\alpha ,~c_\alpha'$ by
\begin{align}
b_\alpha= f_\alpha -\sqrt{\beta},\label{di1s}\\
c_\alpha=c'_\alpha+i\sqrt{\gamma_\alpha}\label{di2s}
\end{align}
one obtains
\begin{align}\label{Y3}
\Pi_\alpha^{\alpha',2} =\sqrt{\omega_{\alpha'}\over 2v}\left(i\left[{c'}_{\alpha'}^\dagger-c'_{\alpha'}\right]+2\sqrt{\gamma_{\alpha'}}\right)-{d\over  v} (\alpha-\alpha')\sqrt{N-\beta}\left(f_{\alpha'}^\dagger \sqrt{\xi_{\alpha'}}+\sqrt{\xi_{\alpha'}} f_{\alpha'}-2\sqrt{\beta \xi_{\alpha'}}\right)
\end{align}
where $\beta,~\gamma_\alpha$ and $\xi_\alpha$ are defined in Eqs.~(\ref{bets}), (\ref{gams}), and (\ref{xi}) respectively. Expanding $\sqrt{\xi_{\alpha'}}$ and retaining terms which do not vanish in the thermodynamic limit yields
\begin{align}\label{Y4}
\Pi_{\alpha,\rm th}^{\alpha',2,{\rm a}} =\sqrt{2\omega_{\alpha'}\gamma_{\alpha'}\over 2v}+2(\alpha-\alpha')d{\sqrt{(N-\beta)\beta}\over v} =\alpha \rho d\sqrt{1-\tau^2},
\end{align}
as given in the main text. We plot $\Pi_{\alpha,\rm th}^{\alpha',2,{\rm a}} $ in Fig.~\ref{momf} for different $\alpha$ again using the example of double-well dipoles. As a special case, this includes the gauge-invariant macroscopic manifestation of the abnormal phase as quantified by $P_{{\rm T} \alpha,\rm th}^{\alpha',2,{\rm a}}$

\section*{
Further numerical results}

\begin{figure}[H]
  \centering
(a)\hspace*{0.3cm}\includegraphics[scale=0.38]{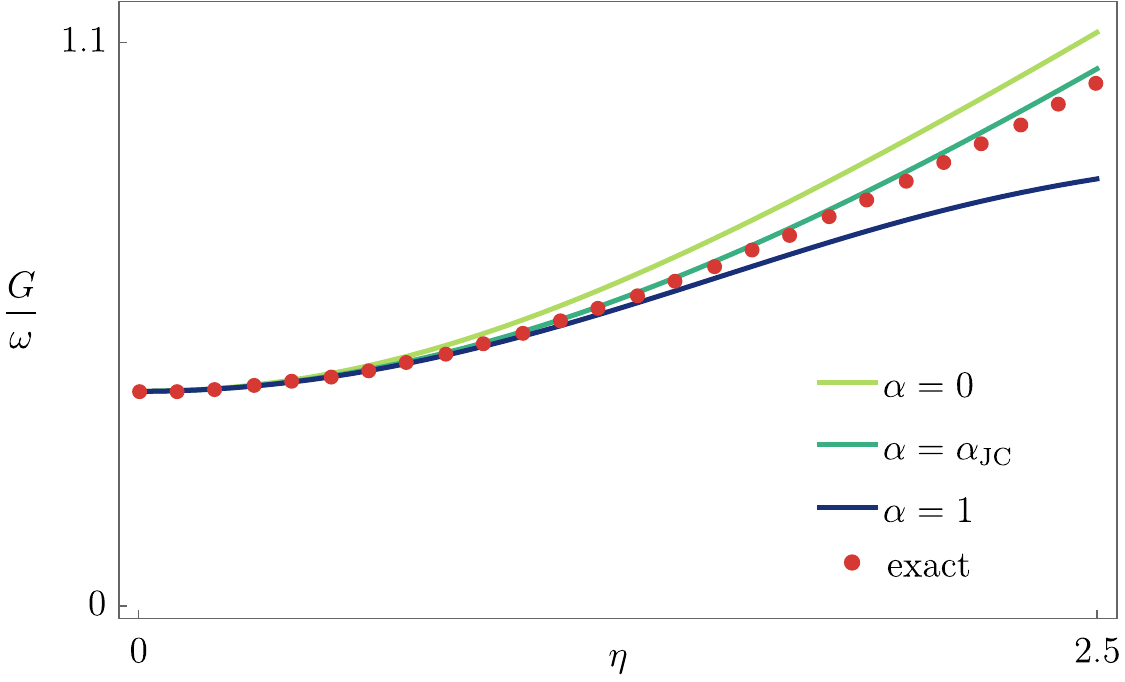}
\vspace*{0.45cm}

\hspace*{-0.2cm}(b)\includegraphics[scale=0.4]{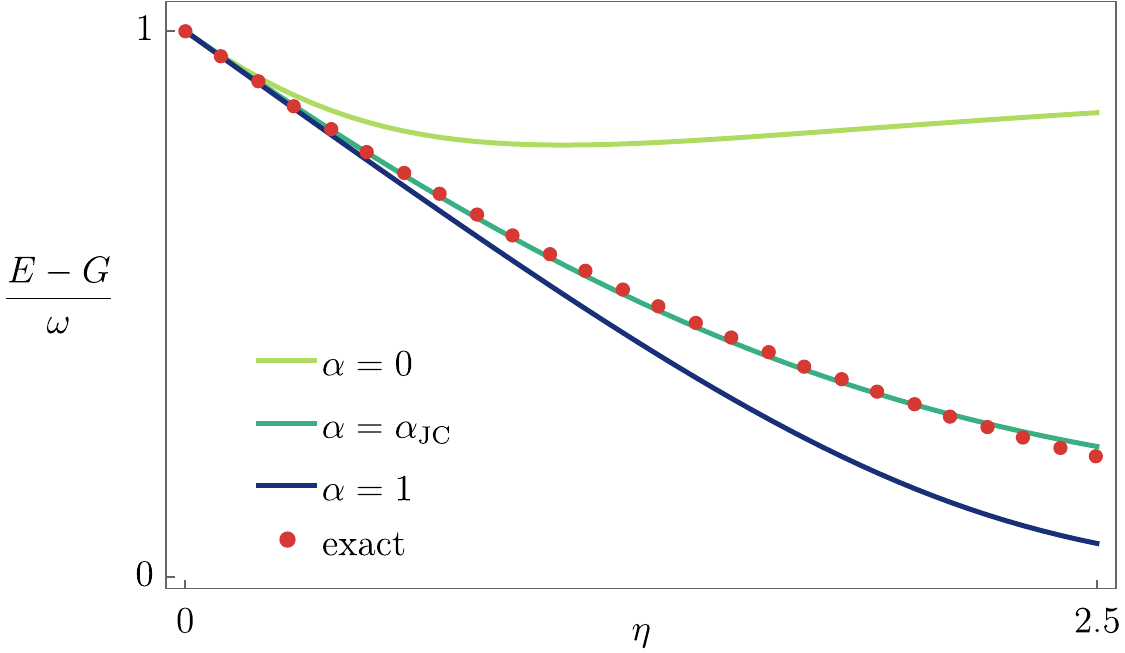}

\hspace*{0.4cm}(c)\hspace*{0.0cm}\includegraphics[scale=0.435]{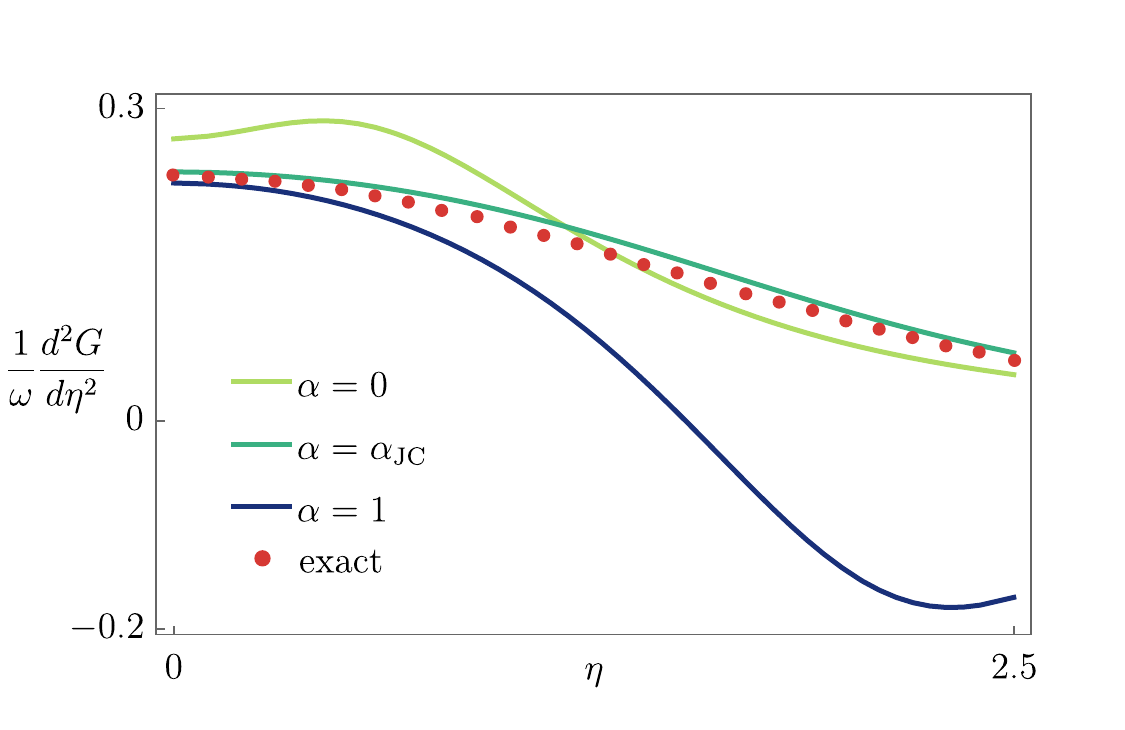}
\caption{In all plots we have chosen $\beta=1.5$ and then chosen ${\cal E}$ such that $\omega_m=\omega$. The single-dipole Coulomb gauge, Jaynes-Cummings gauge and multipolar gauge two-level model predictions are compared with the corresponding exact predictions as a function of $\eta$ for: (a) the ground energy $G$, (b) the first transition energy $E-G$, (c) the second derivative $d^2 G/d\eta^2$. In all cases the Jaynes-Cummings gauge two-level model is most accurate.}
  \label{s1}
\end{figure}

\begin{figure}[H]
  \centering
(a)\hspace*{0.3cm}\includegraphics[scale=0.38]{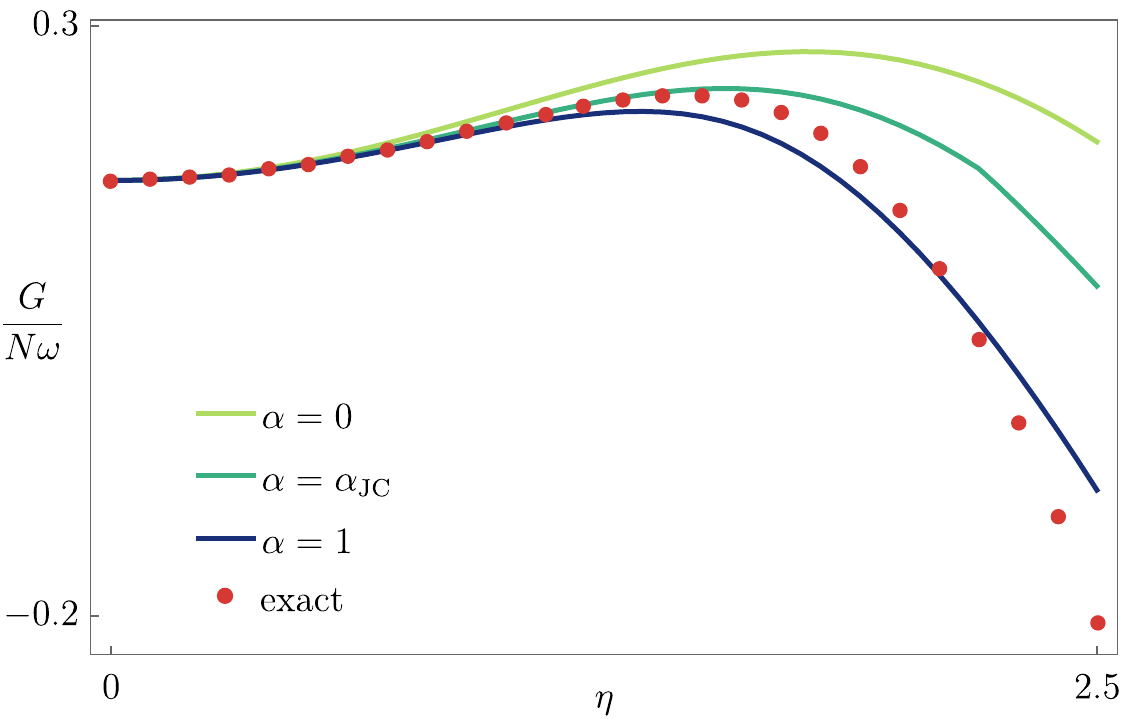}
\vspace*{0.45cm}

\hspace*{-0.2cm}(b)\hspace*{-0.25cm}\includegraphics[scale=0.41]{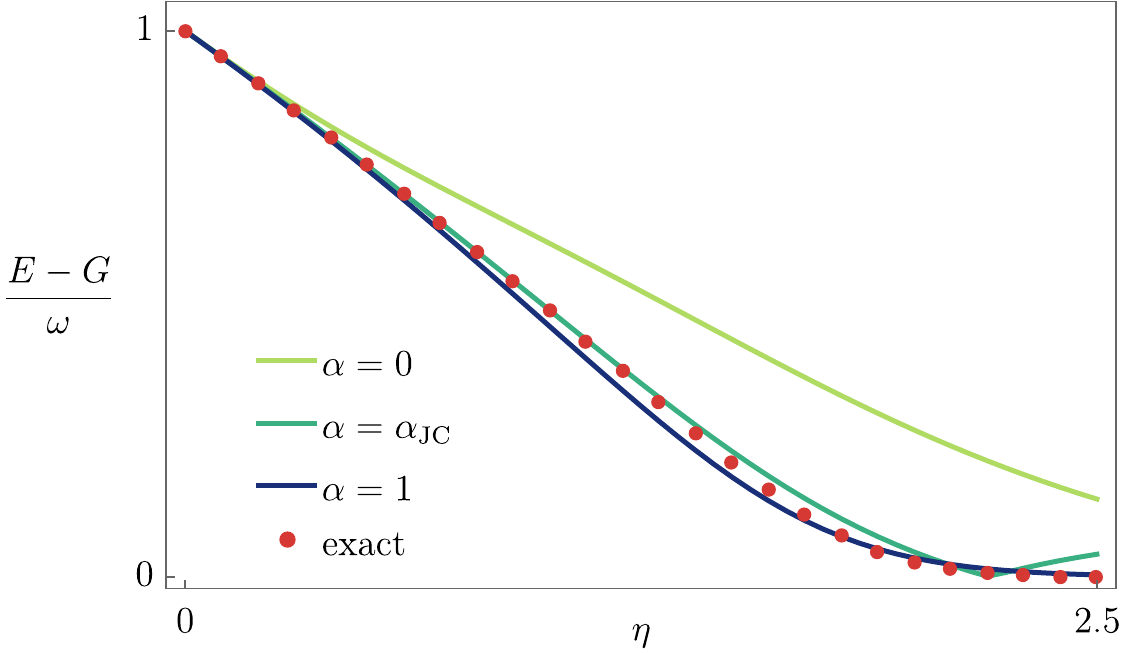}
\caption{In all plots we have chosen $\beta=1.5$ and then chosen ${\cal E}$ such that $\omega_m=\omega$. For the two-dipole case ($N=2$), the Coulomb gauge, Jaynes-Cummings gauge and multipolar gauge two-level model predictions are compared with the corresponding exact predictions as a function of $\eta$ for: (a) the ground energy $G$, (b) the first transition energy $E-G$.}
  \label{s2}
\end{figure}

\begin{figure}[H]
  \centering
(a)\hspace*{0.3cm}\includegraphics[scale=0.38]{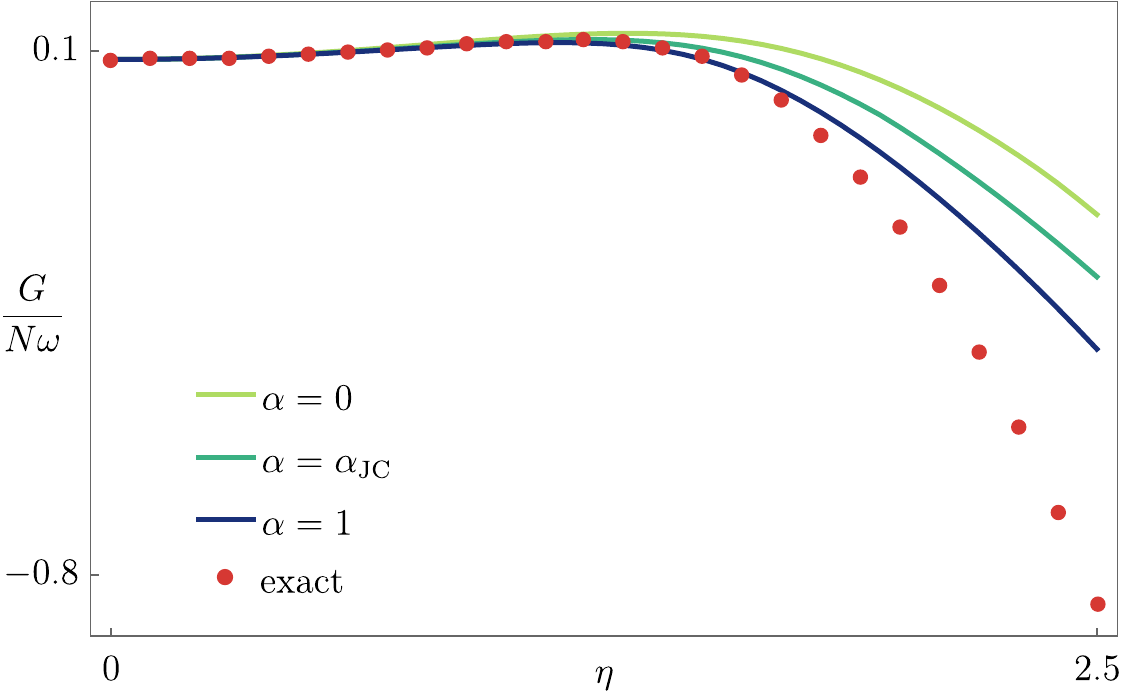}
\vspace*{0.45cm}

\hspace*{-0.2cm}(b)\hspace*{-0.25cm}\includegraphics[scale=0.41]{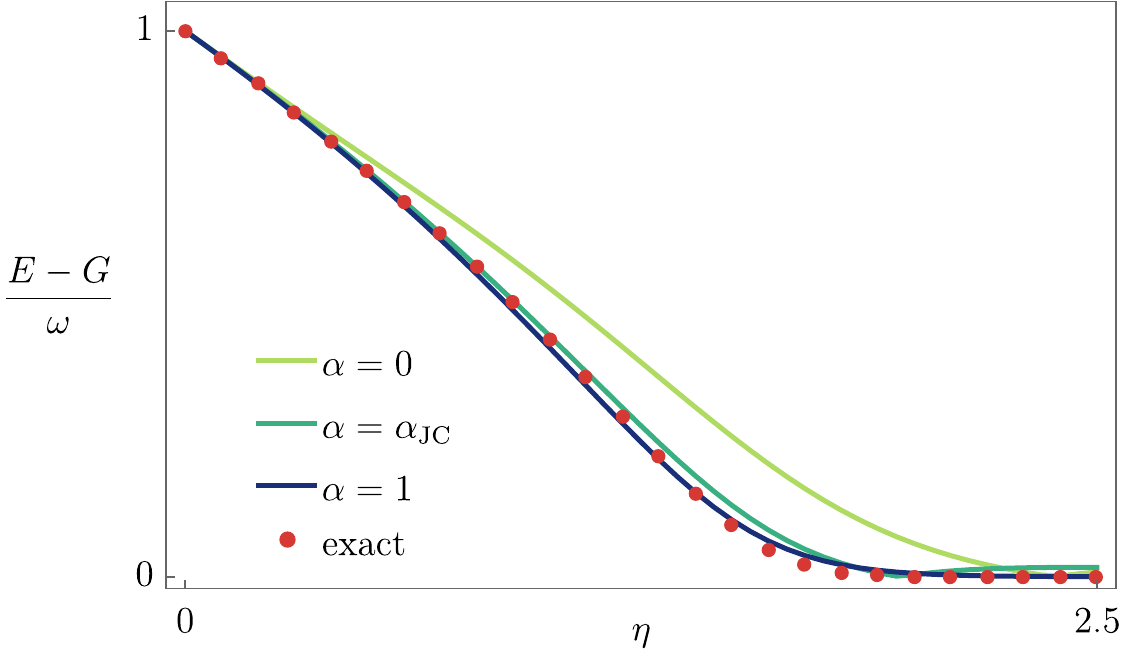}
\caption{In all plots we have chosen $\beta=1.5$ and then chosen ${\cal E}$ such that $\omega_m=\omega$. For the three-dipole case ($N=3$), the Coulomb gauge, Jaynes-Cummings gauge and multipolar gauge two-level model predictions are compared with the corresponding exact predictions as a function of $\eta$ for: (a) the ground energy $G$, (b) the first transition energy $E-G$.}
  \label{s3}
\end{figure}

Here we consider a less anharmonic single-dipole double-well potential, which has $(\epsilon_2-\epsilon_0)/\omega_m\approx 3.2$. This results from choosing $\beta=1.5$ rather than $\beta=3.3$ as was chosen in the main text. In this case single-dipole two-level models are able to remain accurate in predicting the low energy properties of the system, but the multipolar gauge no longer provides the optimal two-level model. The optimal gauge for the two-level truncation is shifted towards the Coulomb gauge, such that the Jaynes-Cummings gauge two-level model is close to optimal \cite{stokes_gauge_2019}. This is shown in Fig. \ref{s1}, which compares $G$ (Fig. \ref{s1}a), $E-G$ (Fig. \ref{s1}b), and $d^2G/d\eta^2$ (Fig. \ref{s1}c) each obtained from the Coulomb-gauge, Jaynes-Cummings gauge, multipolar-gauge two-level models, and the exact (non-truncated) theory. For $N>1$, two-level models become less accurate in predicting even low energy properties when the coupling is sufficiently strong, as shown for the case $N=2$ in Fig. \ref{s2} and for the case $N=3$ in Fig. \ref{s3}. The accuracy of multipolar-gauge two-level truncation appears to improve relative to the other gauges, but the ground energy is not well represented by any two-level model for sufficiently strong coupling.

\end{document}